%% file: main.tex
\documentclass{sig-alternate}
\usepackage{mathptmx} 

\usepackage{fancyhdr}
\usepackage[normalem]{ulem}
\usepackage[hyphens]{url}
\usepackage[sort,nocompress]{cite}
\usepackage[final]{microtype}
\usepackage[keeplastbox]{flushend}
\usepackage{cite}
\usepackage{amsmath,amssymb,amsfonts}
\usepackage{algorithmic}
\usepackage{algorithm}
\usepackage{graphicx}
\usepackage{authblk}
\usepackage{textcomp}
\usepackage{xcolor}
\usepackage{braket}
\usepackage{tikz}
\usetikzlibrary{quantikz}
\usepackage{mdframed}
\usepackage{tabularx}
\usepackage{adjustbox}
\usepackage{rotating}
\usepackage{booktabs}
\usepackage{colortbl}
\usepackage{xcolor}
\usepackage{amsbsy}
\newcommand*\circled[1]{\tikz[baseline=(char.base)]{
            \node[shape=circle,draw,inner sep=1pt] (char) {#1};}}

\definecolor{lightgreen}{HTML}{cdf3df}
\definecolor{lightred}{HTML}{f3bcab}

\usepackage[bookmarks=true,breaklinks=true,letterpaper=true,colorlinks,citecolor=blue,linkcolor=blue,urlcolor=blue]{hyperref}

\makeatletter
\def\@fnsymbol#1{\ensuremath{\ifcase#1\or \P\or \ddagger\or
   \mathsection\or \mathparagraph\or \|\or **\or \dagger\dagger
   \or \ddagger\ddagger \else\@ctrerr\fi}}
\makeatother
    
\title{Microarchitectures for Heterogeneous Superconducting Quantum Computers}
\author[$\ \ $,1, $\dag$]{Samuel Stein\thanks{samuel.stein@pnnl.gov}}
\author[2,$\dag$]{Sara Sussman}
\author[2,7,$\dag$]{Teague Tomesh}
\author[2,$\dag$]{Charles Guinn}
\author[2,$\dag$]{Esin Tureci}
\author[3]{Sophia Fuhui Lin}
\author[2]{Wei Tang}
\author[1]{James~Ang}
\author[4]{Srivatsan Chakram}
\author[1]{Ang Li}
\author[2]{Margaret Martonosi}
\author[3,7]{Fred T. Chong}
\author[2]{Andrew A. Houck}
\author[5]{Isaac L. Chuang}
\author[$\ \ $,5,6]{Michael Austin DeMarco\thanks{mdemarco@bnl.gov}}
\affil[$\dag$]{\textbf{\textit{These authors contributed equally}}}
\affil[1]{Pacific Northwest National Laboratory, Richland, Washington, USA}
\affil[2]{Princeton University, Princeton, New Jersey, USA}
\affil[3]{University of Chicago, Chicago, Illinois, USA}
\affil[4]{Rutgers University, New Brunswick, New Jersey, USA}
\affil[5]{Massachusetts Institute of Technology, Cambridge, Massachusetts, USA}
\affil[6]{Brookhaven National Laboratory, Upton, New York, USA}
\affil[7]{Infleqtion, Chicago, Illinois, USA}

\date{}

\begin{document}
\maketitle
\pagestyle{plain}


\begin{abstract}
Noisy Intermediate-Scale Quantum Computing (NISQ) has dominated headlines in recent years, with the longer-term vision of Fault-Tolerant Quantum Computation (FTQC) offering significant potential albeit at currently intractable resource costs and quantum error correction (QEC) overheads. For problems of interest, FTQC will require millions of physical qubits
with long coherence times, high-fidelity gates, and compact
sizes to surpass classical systems. Just as heterogeneous specialization has offered scaling benefits in classical computing, it is likewise gaining interest in FTQC.  However, systematic use of heterogeneity in either hardware or software elements of FTQC systems remains a serious challenge due to the vast design space and variable physical constraints.

This paper meets the challenge of making heterogeneous FTQC design practical by introducing HetArch, a toolbox for designing heterogeneous quantum systems, and using it to explore heterogeneous design scenarios. Using a hierarchical approach, we successively break quantum algorithms into smaller operations (akin to classical application kernels), thus greatly simplifying the design space and resulting tradeoffs. Specializing to superconducting systems, we then design optimized heterogeneous hardware composed of varied superconducting devices, abstracting physical constraints into design rules that enable devices to be assembled into \emph{standard cells} optimized for specific operations. 
Finally, we provide a heterogeneous design space exploration framework which reduces the simulation burden by a factor of $10^4$ or more and allows us to characterize optimal design points. 
We use these techniques to design superconducting quantum modules for entanglement distillation, error correction, and code teleportation, reducing error rates by $2.6\times$, $10.7\times$, and $3.0\times$ compared to homogeneous systems. 
\end{abstract}

\input{text/introduction}
\input{text/architecture}

\input{text/devices_and_cells}
\input{text/three_archs}

\input{text/conclusion}

\section*{Acknowledgements}
This material is based upon work supported by the U.S. Department of Energy, Office of Science, National Quantum Information Science Research Centers, Co-design Center for Quantum Advantage (C2QA) under contract number DE-SC0012704, (Basic Energy Sciences, PNNL FWP 76274). The Pacific Northwest National Laboratory is operated by Battelle for the U.S. Department of Energy under Contract DE-AC05-76RL01830. This work is funded in part by EPiQC, an NSF Expedition in Computing, under award CCF-1730449; in part by STAQ under award NSF Phy-1818914; in part by the US Department of Energy Office of Advanced Scientific Computing Research, Accelerated Research for Quantum Computing Program; and in part by the NSF Quantum Leap Challenge Institute for Hybrid Quantum Architectures and Networks (NSF Award 2016136) and in part based upon work supported by the U.S. Department of Energy, Office of Science, National Quantum Information Science Research Centers.  F. T. Chong is Chief Scientist for Quantum Software at Infleqtion and an advisor to Quantum Circuits, Inc. S. Sussman is supported by the Department of Defense (DoD) through the National Defense Science \& Engineering Graduate Fellowship (NDSEG) Program. The authors express gratitude to Kaitlin N. Smith, Lev Krayzman, Alec Eickbusch, and Volodymyr Sivak for insightful discussions.

\bibliographystyle{unsrt}
\bibliography{ref,arquin_ref}

\end{document}

%% file: text/introduction.tex
\section{Introduction}

The pursuit of a large-scale fault-tolerant quantum computer (FTQC) that provides a significant advantage over classical computers~\cite{Preskill2021, 1996quant.ph..5011S} has been in progress for nearly three decades. A FTQC would have high-impact applications in cryptography~\cite{Shor1997}, physics~\cite{Feynman1982, byrnes2006simulating}, chemistry~\cite{kassal2011simulating, reiher2017elucidating}, and machine learning~\cite{havlivcek2019supervised, huang2020predicting} with numerous platforms under development. Some of these, particularly trapped ions, neutral atoms, and superconducting qubit systems, have begun to scale.

\begin{figure}[]
    \centering
    \includegraphics[width = 0.65 \columnwidth]{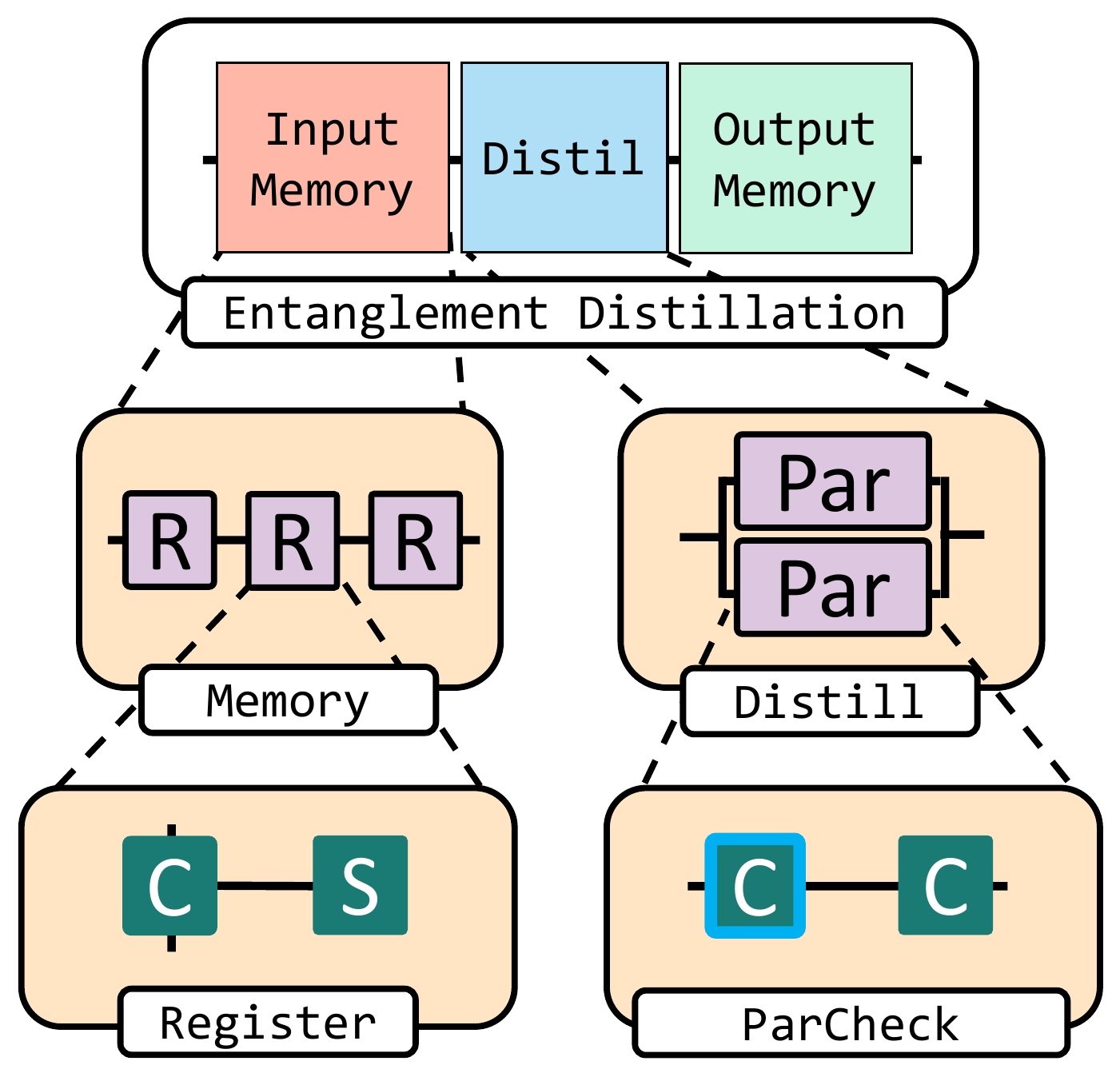}
    \caption{
    Hierarchical module design for entanglement distillation, a protocol for improving entangled pair fidelity. Storage devices (S), such as multimode resonators, provide multi-qubit memory, while compute devices (C), such as transmons, handle routing and two-qubit operations. These are grouped into standard cells and then into submodules for memory and distillation. 
    }
    \label{fig:Distillation_Design}
\end{figure}

The realization of a FTQC has nonetheless been frustrated by high resource costs~\cite{Beverland2022}. Quantum systems are highly sensitive to noise, requiring quantum error correction (QEC)~\cite{PhysRevA.54.1098, nielsen2010quantum}, which introduces an overhead that scales with the error rate of devices. 
In superconducting quantum systems, which are the focus of this paper, 
research efforts have produced a wide variety of interconnectable quantum devices~\cite{2021RvMP...93b5005B} that each trade off between long and short coherence times, slow or fast gates, large or compact sizes, and multi-qubit or single-qubit storage capacity. 
These range from the transmon, a compact 2D qubit with fast gates but shorter coherence times ($O$(500$\mu$s))~\cite{Place2021, Wang2022}) used in the largest systems today~\cite{arute2019quantum, wu2021application} to long-lived resonators that store one or more qubits~\cite{Lei2020, Chakram2021, Ganjam2023MM} with coherence times exceeding 25ms~\cite{2023arXiv230206442M}, but with relatively low connectivity and large sizes. 
However, no platform to date has demonstrated --- simultaneously in a single device --- the required long coherence times, high-fidelity control, compact size, and scale needed for a FTQC, and current estimates for typical applications require hundreds of thousands to millions of physical qubits~\cite{Martinis2015, Beverland2022}. 

On the other hand, a powerful method for reducing the overhead cost of FTQC has been the employment of heterogeneity at both hardware and software levels~\cite{thaker2006quantum}. This has been established for some time at the software level~\cite{metodi2006quantum}, in terms of the heterogeneous operation of qubit arrays, where algorithm implementations~\cite{10.1145/2903150.2906827} which vary the error correcting code or code depth across subroutines have yielded orders-of-magnitude improvements to state distillation~\cite{2019Quant...3..205L} and factoring~\cite{2019arXiv190509749G}. Recent work has also shown the potential of adapting the operation of systems to natural hardware inhomogeneity in coherence times~\cite{2020arXiv201100028S, biasedNoise} and gate sets~\cite{2022arXiv220813380F, 2023arXiv230201252M}. 

Experimental progress has also enabled heterogeneity at the hardware level, where devices can be optimized for specific functions and thus relax simultaneous demands for long coherence times, high-fidelity gates, scalable footprint, and effective topology placed by homogeneous architectures~\cite{Beverland2022}. 
New error correction architectures have been proposed which leverage the differing coherence and gate times of quantum devices~\cite{Hann2023MM, 2020arXiv201204108C, Chamberland2022}. Composite devices consisting of integrated qubit-resonator systems~\cite{2022arXiv221212077T, Chakram2021} have been proposed and realized, with one experiment demonstrating the highest-ever coherence time improvement from QEC~\cite{2023Natur.616...50S}. 

Several recent examples have found further advantages by leveraging software and hardware heterogeneity together. New QEC codes co-designed with novel hardware have found overhead reductions of 3-10$\times$~\cite{Duckering2020, 2022arXiv220504387M} and can implement highly non-planar codes~\cite{2022arXiv220504387M}. An approach to factoring utilizing multimode resonators found an overhead reduction of $1000\times$~\cite{2021PhRvL.127n0503G}. Novel hardware has also inspired quantum simulation approaches~\cite{PhysRevX.10.021060, wangthesis}, with theoretical results indicating cost reduction quadratic in system size~\cite{PhysRevResearch.3.043072}. 

However, despite the success of these ad-hoc examples, the systematic design and validation of heterogeneous quantum systems remains a serious challenge. On the one hand, heterogeneity leads to a combinatorially large design space at each abstraction level that must be navigated, with optimizations interacting across levels~\cite{2019arXiv190509749G}. At the same time, the system must adhere to complex rules at multiple scales, from physical constraints on quantum device connectivity and topology ~\cite{arute2019quantum, PhysRevX.10.011022}, to timing of quantum operations and stochastic processes~\cite{Ang2022}, and these rules depend strongly on design choices of physical hardware and implementation. Compounding this, the complexity of quantum systems prohibits simulation beyond modest sizes~\cite{9355323}, while models for overcoming this limitation rely on analyzing errors~\cite{10.1063/1.1499754} as a system scales, with no clear generalization to situations where errors and their impact vary device-by-device. A framework is needed which simplifies the design process, addressing algorithm requirements, device selection, and ensuring compliance with experimental constraints to enable widespread design of heterogeneous quantum systems.

Of course, classical computing has long employed heterogeneous designs. Modern systems adopt heterogeneity from the level of nodes
~\cite{10.1145/3028687.3038873}, to processors~\cite{tanenbaum1999structured}, down to individual circuit designs~\cite{hall1992microprocessors}. Key to this design process is a hierarchical approach, with microarchitecture~\cite{tanenbaum1999structured} organizing modules inside processors around critical subroutines or application kernels, while Very Large Scale Integration (VLSI) techniques~\cite{1980aw...book.....M, weste1985principles} create physical designs using a hierarchy to break down complex algorithmic needs into successively smaller pieces until a design can be created~\cite{shankar2014vlsi}. However, the adoption of these methods in quantum systems is challenging. Qubit entanglement underlies the advantage of quantum systems~\cite{IBMRoadmap, 2014PhRvA..89b2317M, Ang2022}, but can often work in opposition to the modular approaches familiar in classical computing~\cite{tanenbaum1999structured}. 
In addition, the extreme sensitivity of errors in quantum systems requires error correction, potentially at multiple levels~\cite{2023arXiv230108542T} with varied codes~\cite{1996quant.ph..8012K}. Furthermore, communication in quantum systems admits fundamentally non-classical mechanisms including quantum teleportation~\cite{nielsen2010quantum, 10.1145/1330521.1330522}, given appropriate pre-distribution of Bell pairs~\cite{Ang2022}. 

Still, considerable literature has called for an analogue of large scale design for homogeneous quantum systems ~\cite{9167259, 10.1145/3439706.3446900, 8587760, aspuru-guzik_cad, Kyaw:2020qcz}, with additional focus on the theoretical aspects of device and circuit verification~\cite{8342103, 10.1093/ietfec/e91-a.2.584}. 
At the bottom of the stack, 
libraries such as scQubits~\cite{2021Quant...5..583G, 2022arXiv220608320P} and Qiskit Metal~\cite{Qiskit_Metal} support the detailed layout of the elements of devices used in homogeneous and heterogeneous systems. 
Hybrid systems, consisting of multiple qubit platforms such as ions and atoms or superconducting devices, have been proposed\cite{doi:10.1146/annurev-conmatphys-030212-184253, 2016QuIP...15.5385D, 2012PhRvL.108m0504K}, but transduction challenges~\cite{PhysRevLett.103.043603} have hindered coupling distinct platforms. In the nearer term, overarching visions for modular quantum systems~\cite{2022arXiv220906841B, IBMRoadmap, 2012arXiv1208.0391M, 2019NatCo..10.4692Z, LaRacuente:2022xqq, 9923784} have been proposed, but full-stack heterogeneous design, from software to hardware, remains to be tackled. 

This paper introduces HetArch, a methodology and toolbox for the systematic design and simulation of heterogeneous quantum microarchitectures, and then uses HetArch to develop recommendations for the best uses of heterogeneity in key QC applications. Our contributions include:
\begin{itemize}
\item A \textbf{hierarchical hardware synthesis method} in which modules executing high-level quantum subroutines are recursively broken into basic operations for which heterogeneous physical architectures can be designed.
\item \textbf{Quantum standard cells} as physical architectures optimized for these basic operations, and enumerated \textbf{design rules} to enable their systematic design in compliance with physical constraints while minimizing errors.
\item A heterogeneous \textbf{design space exploration framework} that efficiently simulates performance, even with entanglement, to identify optimal design points among tradeoffs in coherence time, connectivity, and gate sets.
\end{itemize}
We specialize to the particular case of superconducting hardware, where we give an overview of near-term devices and lay out their design rules. We then use this framework to design heterogeneous architectures for \textbf{entanglement distillation, error correction, and code teleportation}, with reductions in error rates of up to $\pmb{2.6\times}$, $\pmb{10.7\times}$, and $\pmb{3.0\times}$ relative to homogeneous systems.

The following section presents an overview of our architectural framework. Section \ref{sec:DevicesAndCells} outlines the available devices and abstracts the physical constraints into architectural design rules to assemble several quantum standard cells. Section \ref{sec:Archs} then presents the three example architectures and demonstrates how they may outperform homogeneous systems. Section \ref{sec:conclusion} discusses how this framework can incorporate other quantum platforms and future experimental progress. 

%% file: text/architecture.tex
\begin{figure*}[th!]
    \centering
    \includegraphics[width=.9\textwidth]{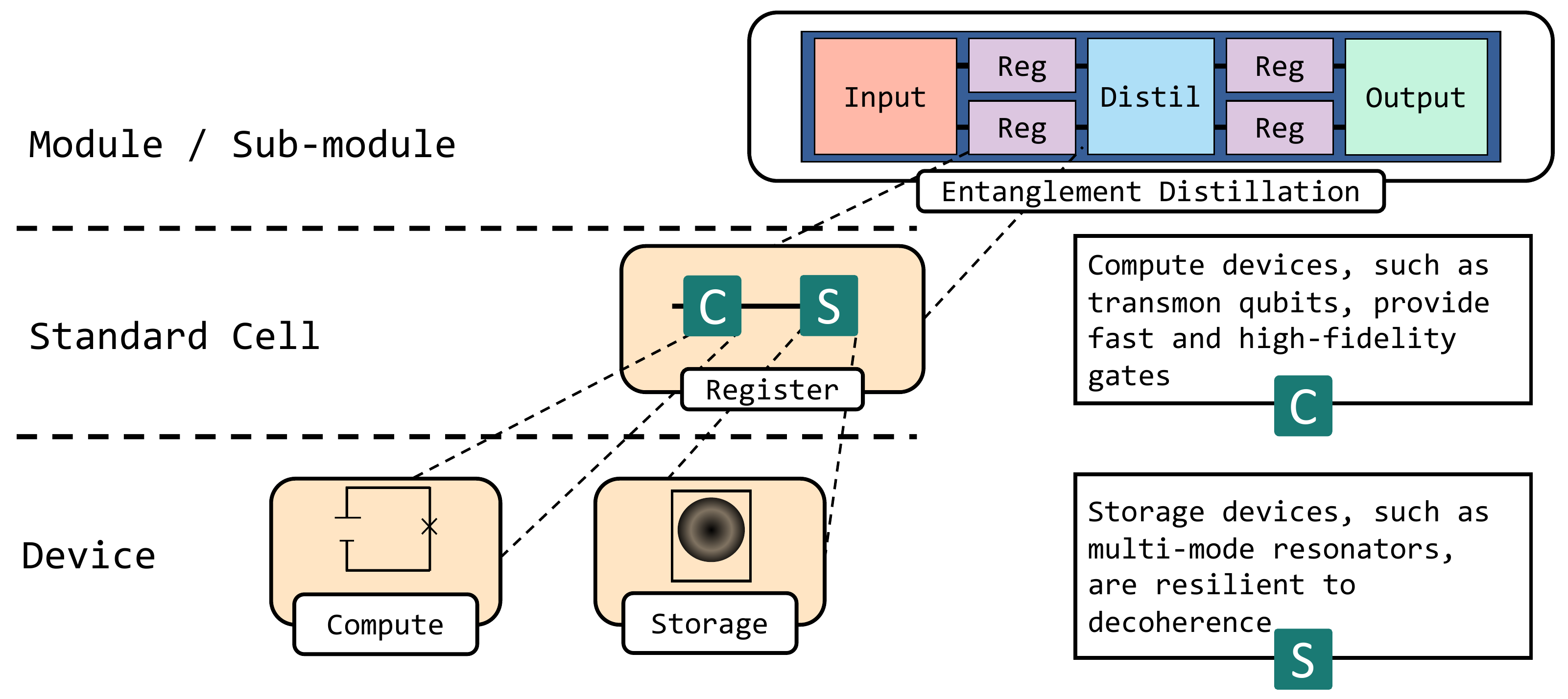}
    \caption{A hierarchical architectural approach that mitigates the exponential complexity of designing heterogeneous systems. High-level quantum subroutines are to be executed by modules, which are designed by breaking the subroutine into successively smaller components until reaching operations sufficiently small that physical architectures can be designed to execute them. This shows the framework for the distillation architecture of Figure \ref{fig:Distillation_Design}.}
    \label{fig:framework}
\end{figure*}

\section{Architectural Framework}\label{sec:Hierarchy}

In contrast to homogeneous ``sea-of-qubit'' architectures, heterogeneous QC architectures differentiate between the various functions that a device may be used for. For example, a device's role can be divided into ``compute,'' when participating in a gate operation, and ``storage,'' when idling. A device optimized for compute functionality requires fast, high-fidelity gates, high connectivity, and a diverse gate set. On the other hand, storage functionality requires long coherence times. By leveraging the natural performance tradeoffs in devices along these lines, a heterogeneous architecture expands the design space for quantum devices.

However, the increased flexibility offered by heterogeneity comes with a price: added complexity in design and operation. Building an effective heterogeneous system necessitates aligning the physical designs and inherent tradeoffs of its components with the performance needs of the subroutines to be executed. These subroutines are intricate, with even the simplest involving tens of `program qubits' that need to be mapped to the physical devices, often placing diverse demands on the lifetimes, gate times, and topology of circuits. As a result, a method is required to break complex subroutines into smaller components that can be more easily managed by architects when specifying a physical design.

As outlined in the introduction, inspiration can be taken from VLSI design principles. A typical VLSI design process~\cite{shankar2014vlsi} begins with high-level functional specification followed by careful design of the logical commands and instruction set of the system, then chip design, and eventually transistor layout. This hierarchical approach allows designers to define key subroutines common to higher-level classical algorithms and break them into smaller subroutines and operations~\cite{6370030}. Once broken into smaller operations, a physical design optimized for each specific operation can be created, called a \emph{standard cell}~\cite{alagic1978design, 1980aw...book.....M, VIJEYAKUMAR20122186}, which complies with \emph{design rules}~\cite{shankar2014vlsi} imposed by the constraints of the physical system and the interconnection needs of modules. Standard cells can be combined into larger blocks, sometimes called modules, until a full system is designed. 
Finally, to verify that the designs perform as expected, simulation is performed in a `simulation hierarchy' \cite{weste1985principles}. Because there is considerable complexity in the design of VLSI systems, the hierarchies are often treated quite flexibly. For instance, different parts of a system may be treated with a different number of levels, these levels may interface in complex ways, and the choice of levels in the hierarchies is a matter of the experience of and convenience for the designers. 

Here we introduce a similar hierarchical architecture for heterogeneous systems. This architecture connects high-level quantum subroutines that are offered as fundamental operations to a user, such as state distillation, error correction, etc., with the low-level physical implementation. At the lowest level, such an architecture can be mapped to an exact physical layout using existing frameworks such as Qiskit Metal~\cite{QiskitMetal}. The goal of this framework is to translate from the needs of a specific quantum subroutine to an abstract layout that can be denoted symbolically in terms of devices, with design rules constraining the resulting physical layout into one that could yield a high-fidelity system.


Figure \ref{fig:framework} shows a high-level overview of our proposed framework for a hierarchical design of heterogeneous quantum computing architectures. Specifically, there are three layers of abstraction, namely \emph{modules}, \emph{standard cells (`cells')}, and \emph{devices}:
\begin{itemize}
\itemsep0em 
    \item \emph{Modules} are responsible for executing \emph{subroutines} for quantum algorithms. Each module offers a specified set of operations on specific input states (if any). These operations are characterized in terms of the average execution time, logical error rate (or fidelity for non-error-corrected systems), and concurrency of operations. Note that I/O is often a key operation for performance~\cite{Ang2022}. Physically, the module inherits a control overhead and physical footprint from the layers below. 
    \item \emph{Standard cells} perform \emph{operations} such as sequential entangling gates, readout, or syndrome operations. They are the elements which are typically tiled to scale up a system. Examples include a small memory register element, the unit cell of an error correcting code, or routing elements. Standard cells offer a specified set of operations, and are characterized by detailed simulation of these operations, extracting the time, fidelity, and concurrency. Standard cells inherit their control overhead and footprint from their constituent devices. 
    \item \emph{Devices} are the fundamental physical elements capable of storing and manipulating quantum information, such as transmon qubits or multimode resonators. Each device offers a wide array of potential gates which can act on arbitrary input states. These gate operations are characterized by their speed and fidelity, while devices also should be labeled by their control overhead and footprint.
\end{itemize}
These elements form parallel software and hardware hierarchies: modules execute subroutines, cells execute operations, and devices hold qubits. While in VLSI the hierarchies for software, hardware, and simulation may differ due to system complexity, in these simpler quantum systems the three will be coincident. Furthermore, the hierarchy for systems and their interfaces should be flexible, with modules potentially becoming sub-modules, and standard cells appearing as sub-cells. The labels of module, standard cell, and device are guides to the operations performed and the level at which each of these layers is characterized. Devices, being the atomic unit, cannot be sub-devices. Note the use of standard cell at the physical level, in contrast to recent approaches at the error correction level~\cite{2022arXiv220604990D}.

The initial step in designing a physical system involves breaking down a quantum algorithm into progressively smaller parts: subroutines executed by modules or submodules and operations executed by standard cells or subcells. This process may result in multiple levels of modules, submodules, cells, and subcells. Care should be taken to ensure that modules reflect underlying device tradeoffs, such as grouping compute and memory-intensive operations.

Upon reaching the lowest level of cells, the specific operations should be simple enough for design. This fundamental step entails selecting devices and assembling them while adhering to physical constraints abstracted into design rules. Multiple drafts of a standard cell may be designed and their performances compared before settling on a final candidate.

The last step is performance evaluation, which must be hierarchical to mitigate exponential complexity and prevent large scale entanglement from disrupting abstraction layers. At the standard cell level, detailed density matrix computations characterize fidelity and execution time. To characterize a module, multiple standard cells can be jointly simulated by exchanging density matrices. At higher levels, the final output of a module includes execution time and a logical error rate (or fidelity for non-error-corrected operations) relative to the intended output state. To mitigate exponentially growing simulation cost, the performance of several modules is modeled through phenomenological error analysis~\cite{10.1063/1.1499754}, evaluated in comparison to expected input and output states. This ensures that simulation performance remains efficient when coupling modules together.

%% file: text/devices_and_cells.tex
\begin{table*}[t]
\centering
\small
 \resizebox{\textwidth}{!}
    {
    \begin{tabular*}
    {0.78\textheight}{>{\raggedright\arraybackslash}p{0.13\textheight}>{\raggedright\arraybackslash}p{0.07\textheight}>{\raggedright\arraybackslash}p{0.04\textheight}>{\raggedright\arraybackslash}p{0.04\textheight}>{\raggedright\arraybackslash}p{0.08\textheight}>{\raggedright\arraybackslash}p{0.06\textheight}>
    {\raggedright\arraybackslash}p{0.06\textheight}>
    {\raggedright\arraybackslash}p{0.08\textheight}>{\raggedright\arraybackslash}p{0.12\textheight}}
       \textbf{Device} &  \textbf{T$_1$/T$_2$} & \textbf{Readout time} & \textbf{Gate set} & \textbf{Gate error (time)} & \textbf{Connectivity} & \textbf{Control Overhead} & \textbf{Footprint} & \textbf{Notes} \\ \hline
        Fixed frequency qubit \cite{Wang2022,Wei2022} & 300$\mu$s / 550$\mu$s & 1$\mu$s & Arb. 1Q/2Q & 1e-3 (100ns) & 4 & 1 charge\linebreak 1 readout & 2 mm x 2 mm & e.g. Transmon\\ \hline
       Flux tunable qubit \cite{Dogan2022, ding2023highfidelity} & 800$\mu$s / 200$\mu$s & 1$\mu$s & Arb. 1Q/2Q & 1e-3 (100ns) & 4 & 1 charge\linebreak 1 flux\linebreak 1 readout & 2 mm x 2 mm & e.g. Fluxonium\\ \hline

       3D quantum memory \cite{Milul2023,Burkhart2021} & 25ms / 30ms & N/A  & SWAP & 1e-2 (1$\mu$s)  & 1 & N/A & 50 mm x 0.5 mm x 1 mm & Requires 2D/3D integration
       \\ \hline

       3D multimode resonator (10 modes)\cite{Chakram2021}  & 2ms / 2.5ms & N/A & SWAP & 1e-2 (400ns)  & 1 & N/A & 100 mm x 100 mm x 10 mm & Requires 2D/3D integration 
       \\ \hline
       
       Future on-chip multimode resonator \cite{Leung2019,Chakram2021,Ganjam2023MM} & 1ms / 1ms & N/A & SWAP
       & 1e-2 (100ns) & 1 & N/A & 5 mm x 5 mm  & No demonstration \\ \hline
     \end{tabular*}}
          \caption{Properties of near-term superconducting quantum devices (values estimated from Device column references). The best observed properties for each device were reported and these values have not been demonstrated at scale. For a discussion of the near-term viability of demonstrating a 1ms on-chip multimode resonator, see Section \ref{sec:Devices}.}
     \label{tab:DeviceTable1}
 \end{table*}

\section{Devices and Standard Cells}\label{sec:DevicesAndCells}
Once an algorithm has been broken down into basic operations, design then proceeds `bottom-up,' assembling devices into standard cells optimized for those basic operations, with standard cells then grouping into modules. 
\subsection{Quantum Devices}\label{sec:Devices}


Devices are the fundamental physical objects used for quantum information processing. Various superconducting devices have been created, ranging from compact $500 \mu$m x $500 \mu$m 2D transmon qubits with fast gates to 3D resonators with multiple-qubit storage capacity and a size of $100$cm$^3$. Here we outline these devices, how to characterize them, and how we can group them for use in a heterogeneous system. 

A central tradeoff in superconducting quantum devices is the competition between long coherence times required for quantum information storage and high connectivity desired for computation~\cite{Gao2021}. This tradeoff is present both within a type of device (for example, optimizing gate speed may require drive lines with higher couplings that lower coherence time), as well as between device types (as we will see for the longer coherence times and reduced connectivity of 3D resonators). Increasing coherence times while preserving connectivity is an ongoing engineering challenge.

Heterogeneity adapts the system to this tradeoff by choosing devices based on software needs~\cite{Thaker2006}. In this initial study, we group devices into `compute' and `storage' functions, which can then be mapped onto the demands of quantum circuits. Compute devices have high connectivity and fast, high-fidelity gates, with single-qubit capacity. Storage devices have low connectivity, to preserve long coherence times, and multi-qubit capacity. Additional tradeoffs, including potentially new technologies, are discussed in Section \ref{sec:conclusion}.

Table~\ref{tab:DeviceTable1} shows the key properties of devices. The coherence times $T_1$ and $T_2$ define the timescale for amplitude and phase damping errors, respectively. A device's readout time is the time required to measure the system (e.g. in the $Z$ basis). Planar devices (transmons, etc.) are measured by coupling to readout circuitry while resonators are measured by coupling to a qubit and then readout. A device's gates are characterized by their typical durations and average gate fidelities. The connectivity of a device is the number of connections allowed. For the case of resonators, a single transmon is connected to the resonator which can then be connected to other devices. Control overhead is a measure of the extra I/O required to operate a device. For example, the fluxonium can achieve higher $T_1$ than a fixed frequency transmon but requires a dedicated flux bias line \cite{Nguyen2019}.

The primary compute device considered in this work is the planar transmon. The transmon is the only superconducting qubit for which high-fidelity gates have been successfully scaled up to a processor with over two devices \cite{Krinner2022,Acharya2023}. The transmon has known coherence limitations~\cite{Read2022, Gao2021, Place2022}, with the highest demonstrated transmon coherence times currently well below 1ms \cite{Wang2022}. While the transmon can be made with a tiny footprint of 1$\mu$m x 1$\mu$m \cite{PhysRevApplied.16.024023, Mamin2022MM}, this design has not yet been scaled up to a multi-qubit system.

The primary storage device considered here is the multimode resonator, which functions as a small multi-qubit quantum memory~\cite{Chakram2021}. Resonators have already been experimentally implemented as quantum memories that store the state of planar superconducting qubits \cite{Mariantoni2011, Reagor2016}. Experimental demonstrations of high-coherence 3D multimode resonators showed coherence times of over 2ms for 8 modes accessible via a single transmon that can store and load qubits with 95\% fidelity, 400ns long \texttt{SWAP} gates exchanging states between the transmon and resonator~\cite{Chakram2021}.  Moreover, the gate fidelity is expected to be limited only by \texttt{SWAP} gate time and the $T_2$ of the transmon, so further prototyping should result in fast, high-fidelity gates similar to those between two transmons. Experimental efforts to develop efficient multimode planar resonator designs are underway \cite{Matanin2023}.

For ultra-high coherence storage, 3D resonators are a promising candidate with coherence times as high as 25ms~\cite{Milul2023}, but these have a large footprint and will be more challenging to precisely couple to 2D devices~\cite{Axline2016, Burkhart2021}. New experimental demonstrations allow us to consider storage devices that could be integrated on-chip. Single-mode planar resonators can now have coherence times of 1ms~\cite{Ganjam2023MM}, and micromachined resonators can have coherence times of 5ms~\cite{Lei2020}.  A future option may be nanomechanical resonators with coherence times exceeding 1 second~\cite{MacCabe2020} if they can be coupled to supercomputing qubits~\cite{Pechal2019}. 

The compute-storage assignment is a major simplification of the full range of quantum devices, but in later sections will suffice to enable a wide range of new architectures. Importantly, a single storage device provides both long coherence times to and also many-to-one connectivity between the qubits stored in it and an attached compute device, akin to set associative access in a classical cache through a single port. Furthermore, different qubit platforms will likely have different tradeoffs, and thus require a different approach to grouping devices, which we will return to in Section~\ref{sec:conclusion}.




\subsection{Quantum Standard Cells}\label{sec:StandardCells}



Standard cells are functional units built from devices and optimized to perform a few quantum operations. Standard cells form the building blocks of a functional module or sub-module. However, standard cells must be assembled in compliance with physical design constraints, such as connectivity and available operations, in a way which best enables the operations they are to support. Here we list design rules which ensure that a standard cell can be physically implemented, and show how to craft several simple standard cells. 



The physical constraints that systems must obey arise from both footprint constraints as well as the need to maximize coherence \cite{Gao2021,2021RvMP...93b5005B}. The complex series of drive lines and couplings~\cite{Krinner2019,Bravyi2022} that are used to manipulate quantum information have a large footprint~\cite{Krinner2022}, with flip-chip architectures only somewhat alleviating 2D space constraints~\cite{Brecht2016, Kosen2022}. On the other hand, every coupling creates a potential vector for quantum information to leak, thus harming coherence~\cite{Wang2022}. In particular, qubits with readout capabilities are expected to have lower coherence~\cite{Sete2015}. 
These considerations lead to the following design rules (DRs) for planar devices:
\begin{mdframed}
\begin{enumerate}
    \item \textbf{Compute devices should be connected to at most 4 other devices.}
    \item \textbf{Storage devices should be connected to exactly 1 compute device to maximize coherence.}
    \item \textbf{Compute devices should have the minimal number of connections required without introducing extra SWAPs.}
    \item \textbf{Compute devices with readout capabilities should be minimal without introducing extra SWAPs.} 
\end{enumerate}
\end{mdframed}

The performance of a given standard cell is characterized through density matrix simulations at the device level, yielding an output density matrix that can be used to extract the relevant metrics such as operation fidelity. Fidelity computed as error rates and coherence are then used to model each standard cell as a quantum channel on its inputs/outputs, abstracting away all of the device level details, which is key to the scalability of the HetArch methodology. Note that, since standard cells exist one layer above the Device layer, it is possible to swap out different physical devices in a standard cell, e.g. changing the choice of storage unit. Doing so will change the performance of the standard cell as well as its footprint and control overhead.

\begin{table*}[]
    \centering
    \setlength{\fboxsep}{7pt}
    \fbox{
    \begin{tabularx}{0.95\linewidth}{l X l}
       & \adjustbox{valign=c}{\includegraphics[width=2.4 cm]{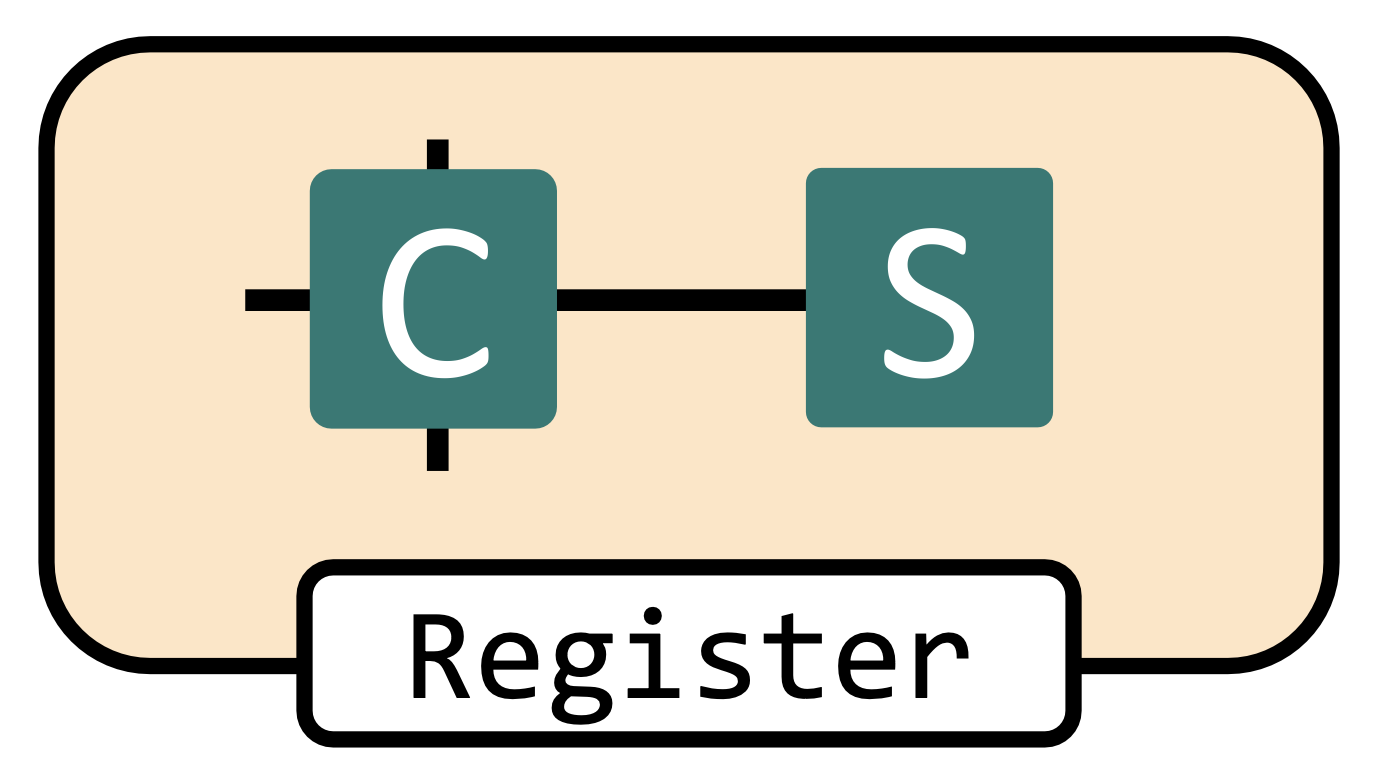}}  \hspace{1cm}\parbox[c]{\dimexpr\linewidth-2\tabcolsep - 3cm}{ \textbf{Register standard cell} \texttt{Register}: A high-capacity storage device coupled to a compute device which manages input/output, with up to three connections from the compute device. Characterized by the load/save time and fidelity to swap a qubit between compute and storage as well as the storage decay time $T_S$.} &  \\ \addlinespace[10pt]
       & \adjustbox{valign=c}{\includegraphics[width=2.4 cm]{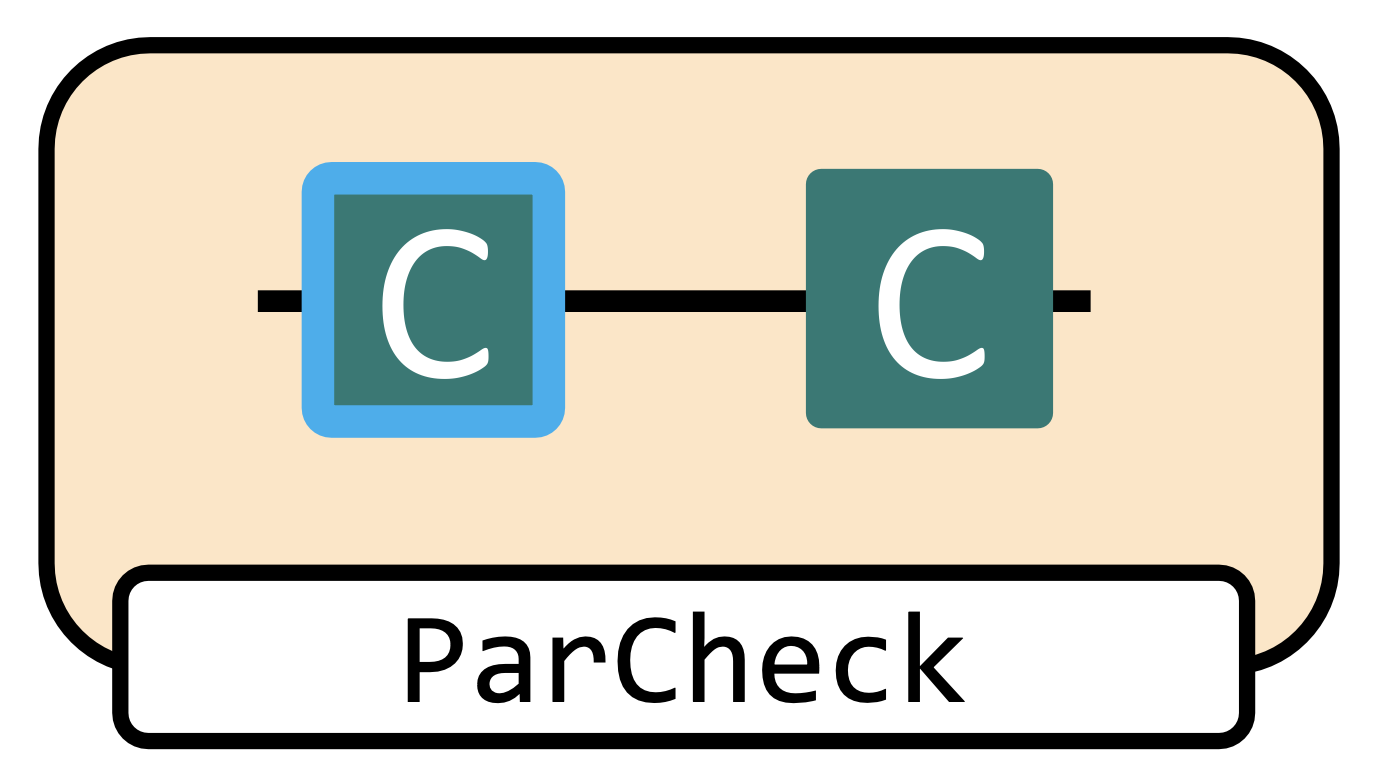}}  \hspace{1cm}\parbox[c]{\dimexpr\linewidth-2\tabcolsep - 3cm}{\textbf{Parity Check Cell} \texttt{ParCheck}: Two compute devices coupled to read data in, perform one and two-qubit operations, and then measure one qubit, with up to three connections from each qubit. Characterized time and fidelity to move two qubits in and out, one and two-qubit gates, and readout time.}  & \\
       \addlinespace[10pt]
       & \adjustbox{valign=c}{\includegraphics[width=2.4 cm]{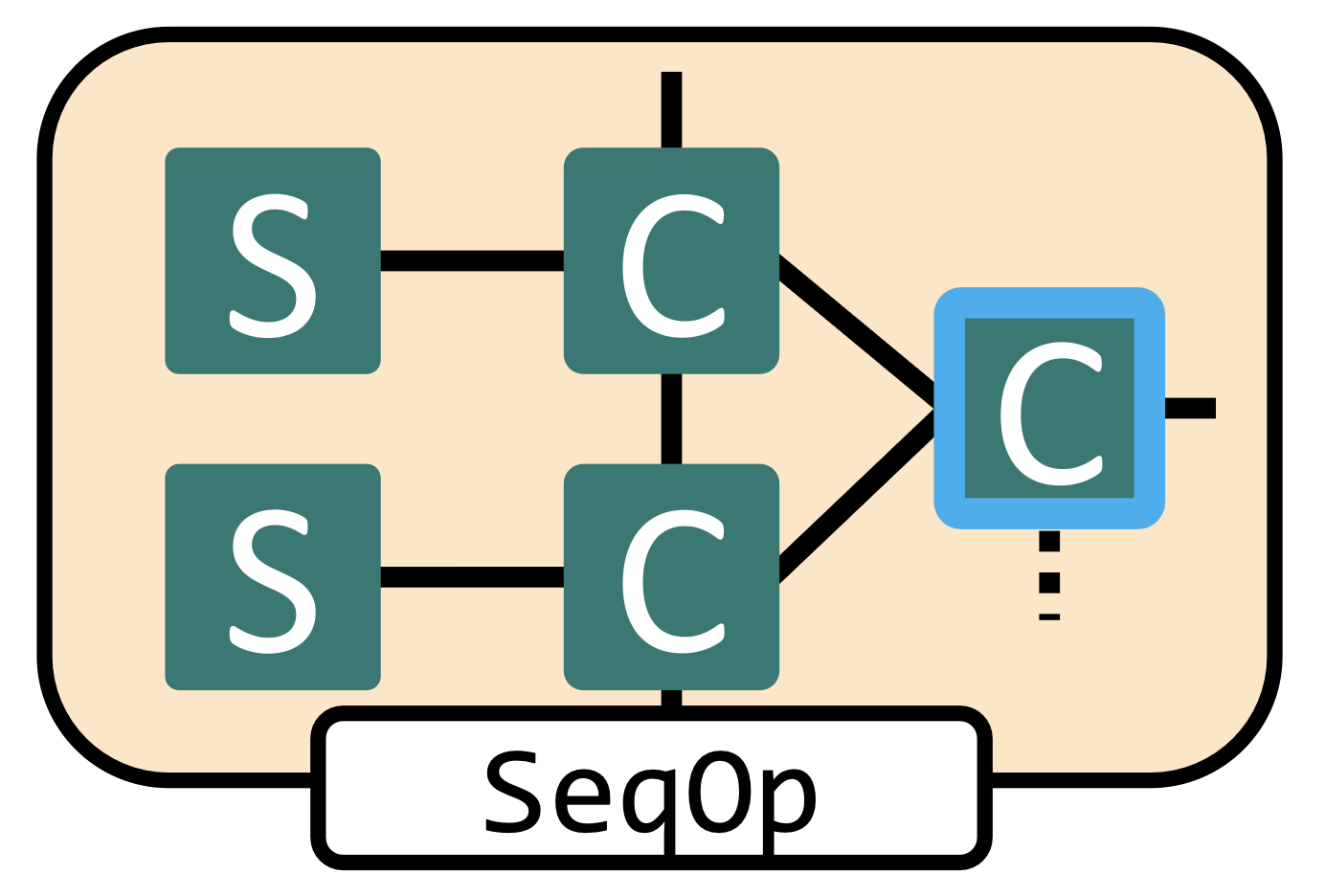}}  \hspace{1cm}\parbox[c]{\dimexpr\linewidth-2\tabcolsep - 3cm}{\textbf{Sequential Operations Cell} \texttt{SeqOp}: Optimized for many sequential two-qubit operations and parity checks among a collection of qubits. Contains two \texttt{Register} standard cells as subcells, coupled to each other and a compute device with readout for parity checking. There are up to two connections from each \texttt{Register} compute device and an optional connection from the parity check compute. Characterized by the time and fidelity to execute a series of two-qubit gates among qubits stored in the \texttt{Register} subcells.} & \\
       \addlinespace[10pt]
       & \adjustbox{valign=c}{\includegraphics[width=2.4 cm]{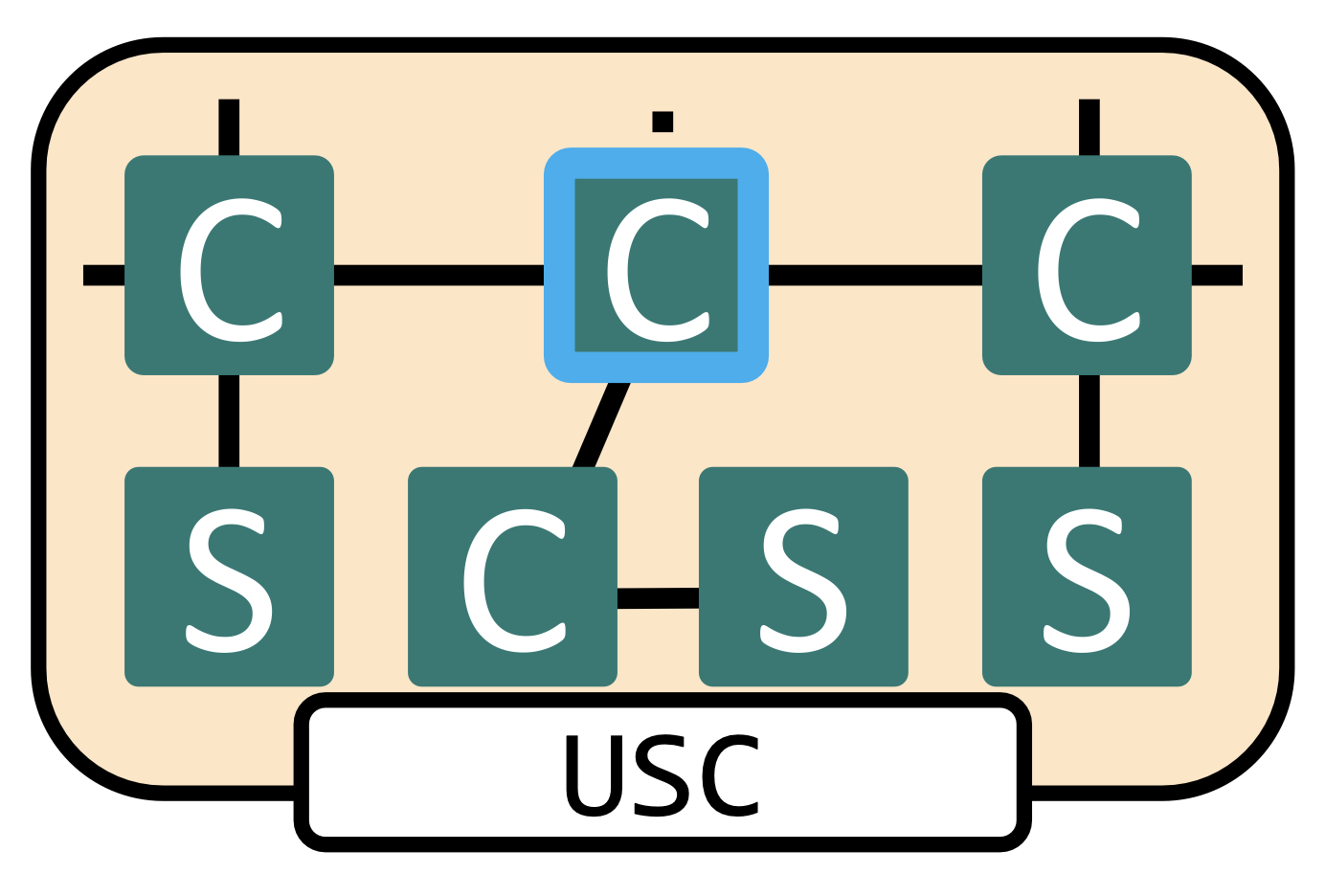}}  \hspace{1cm}\parbox[c]{\dimexpr\linewidth-2\tabcolsep - 3cm}{\textbf{Universal Stabilizer Cell} \texttt{USC}: Contains three \texttt{Register} standard cells as subcells, with a central parity check compute device, with one connection available from each \texttt{Register} compute device and the parity check compute device. Characterized by the time and fidelity to execute stabilizer checks among qubits stored in the \texttt{Register} subcells, with the central parity check device holding the ancilla qubit.} &
    \end{tabularx}
    }
    \caption{Quantum Standard Cells used in this paper. These are assembled in accordance with the design rules presented in Section \ref{sec:Devices}, and optimized for particular operations. Devices with readout are outlined in blue.}
    \label{tab:StandardCellTable}
\end{table*}

To illustrate the process of designing a standard cell, we now examine two which will be used in what follows: a register cell, \texttt{Register}, and a cell optimized for a parity check operation needed for distillation, \texttt{ParCheck}. Table~\ref{tab:StandardCellTable} shows all four standard cells used in this paper, with {\tt SeqOp} and {\tt USC} described as they are used in subsequent sections.





    

The register cell \texttt{Register}, shown in Table~\ref{tab:StandardCellTable} is designed to minimize errors during idle time while allowing high-fidelity movement to and from the compute device. It should accept incoming qubits, store them, and release them out as needed. To do so, it uses a storage element that is coupled to a single compute element (DR2), without readout (DR4). The compute device may be connected to up to three other devices (DR1), though the minimal number should be used (DR3).


A second simple standard cell that we will use is the \texttt{ParCheck}, a cell optimized for parity checks, which must be capable of doing single and two-qubit gates and qubit readout. It is made of two compute devices optimized for fast single-qubit gates and a two-qubit gate between them. One device has a readout resonator allowing parity checks (DR4). It has one connection between the compute devices, allowing each end to link with up to three other units (DR1, DR3).

%% file: text/three_archs.tex
\section{Three Heterogeneous Microarchitectures}\label{sec:Archs}

The architecture presented in Section \ref{sec:Hierarchy} provides an overview of the heterogeneous design methodology, with the devices and standard cells following the constraints laid out in the design rules. We now employ the devices and standard cells of Section~\ref{sec:DevicesAndCells} to demonstrate the design process and evaluation of heterogeneous microarchitectures for three examples: entanglement distillation, error correction, and code teleportation, which leverage the long lifetimes of storage units, heterogeneity in compute units, and topology enhancements offered by heterogeneous design. We sweep design parameters of each and quantify the performance improvements in the heterogeneous systems. 

To compare each heterogeneous architecture, consisting of compute and storage devices, fairly to the to the homogeneous case, we set as a baseline a system consisting of only compute devices arranged in a square lattice, as these underlie many `software-only' heterogeneous approaches~\cite{2019arXiv190509749G, 2019Quant...3..205L}. While the homogeneous system lacks the long-lived, high capacity storage units, it is allowed to be as large as needed for maximally efficient transpilation, with the caveat that error correction will only be applied if it is applied in the heterogeneous system.
If an optimal square lattice transpilation is known, as in the case of surface code, it will be used; otherwise the Qiskit transpiler~\cite{Qiskit} at the highest optimization setting is used. 

Throughout this section, compute and storage coherence times are denoted as $T_c$ and $T_s$, with $T_1$ = $T_2$. Unless otherwise specified, compute devices will have lifetimes $T_C= 500\mu s$, while the wide array of storage options allows $T_S$ to vary from $500\mu s$ to $50 $ms. All gates are assumed to be coherence-limited, with two-qubit gate times (including \texttt{SWAP}s) of 100ns, single qubit gate times of 40ns, and 1$\mu$s error-free readout. Classical communication times are neglected.

\input{text/distillation}

\input{text/memory}
\input{text/teleportation}

%% file: text/distillation.tex
\subsection{Entanglement Distillation}\label{sec:entanglement-distillation}

The production and distribution of entanglement is a key subroutine within many quantum applications. This is typically achieved by preparing and distributing a Bell state entangled pair (EP), for example $\frac{1}{\sqrt{2}}(\ket{00} + \ket{11})$. Current EP generation methods, including on-chip distribution, off-chip microwave connections, and hybrid microwave-optical schemes, suffer from noise and slow generation rates. Entanglement distillation protocols can correct noise by consuming multiple low-fidelity EPs to create a smaller number of higher fidelity EPs. Distillation protocols are expected to play a key rule in networked quantum systems~\cite{PhysRevLett.127.040503}, and here we focus on the DEJMPS protocol~\cite{DEJMPS_1996}. 

However, the effectiveness of distillation is severely impaired by low EP generation rates, as errors accumulate in stored EPs awaiting sufficient numbers for distillation~\cite{Ang2022}. 
Heterogeneous design can alleviate this by leveraging the differing lifetimes of compute and storage components to provide both a memory for storing EPs and a fast distillation protocol. Here, we design a module for entanglement distillation which purifies EPs until they reach a threshold fidelity and then offers them on demand to the rest of the system. We use our quantum standard cells, and focus on hardware parameters comparable to the case of microwave-to-optical conversion \cite{Ang2022}: target fidelity is set to $99.5\%$ and EP generation is random, with average times of 1-100$\mu$s and infidelities on the order of .01-.1, 10-1000x slower and 10-100x noisier than compute operations.

The distillation operation comprises three subroutines. First, EPs are preserved in an input memory until enough are acquired for distillation. Next, the distillation protocol, including gates, measurement, and correction, is executed. Finally, EPs are preserved in an output memory until they are used by other modules. The entanglement distillation module contains a submodule for each of these steps.  

An example entanglement distillation module is shown in Figure~\ref{fig:Distillation_Design}. The input memory consists of one or more \texttt{Register} standard cells. The distillation submodule is comprised of \texttt{ParCheck} standard cells, which can perform the parity checks needed for distillation. Finally, the output memory is also a collection of \texttt{Register} standard cells. In addition to the physical design of the distillation module, a complication arises in the design of its operation and scheduling. Because we assume the case of probabilistic EP generation, the distillation module must dynamically respond to successful generations requiring complex coordination. We implement a greedy scheduler with the following priorities: (1) re-distill existing pairs if it would yield improvement, (2) move distilled pairs to output memory (3) distill new pairs if available (4) store incoming pairs in memory. 

Capacity demands for memories and the distillation module are determined by input EP rates, distillation times, desired output rates, and bottlenecks between these factors. Sweeping the parameter space with a single EP input line with generation rate of $.1-100$MHz, two \texttt{Register} cells for the input memory, one \texttt{ParCheck} cell for distillation, and one output \texttt{Register} were found sufficient to achieve high fidelity distilled EPs without overflow in any submodule.

\begin{figure}[]
    \centering
    \includegraphics[width=\columnwidth]{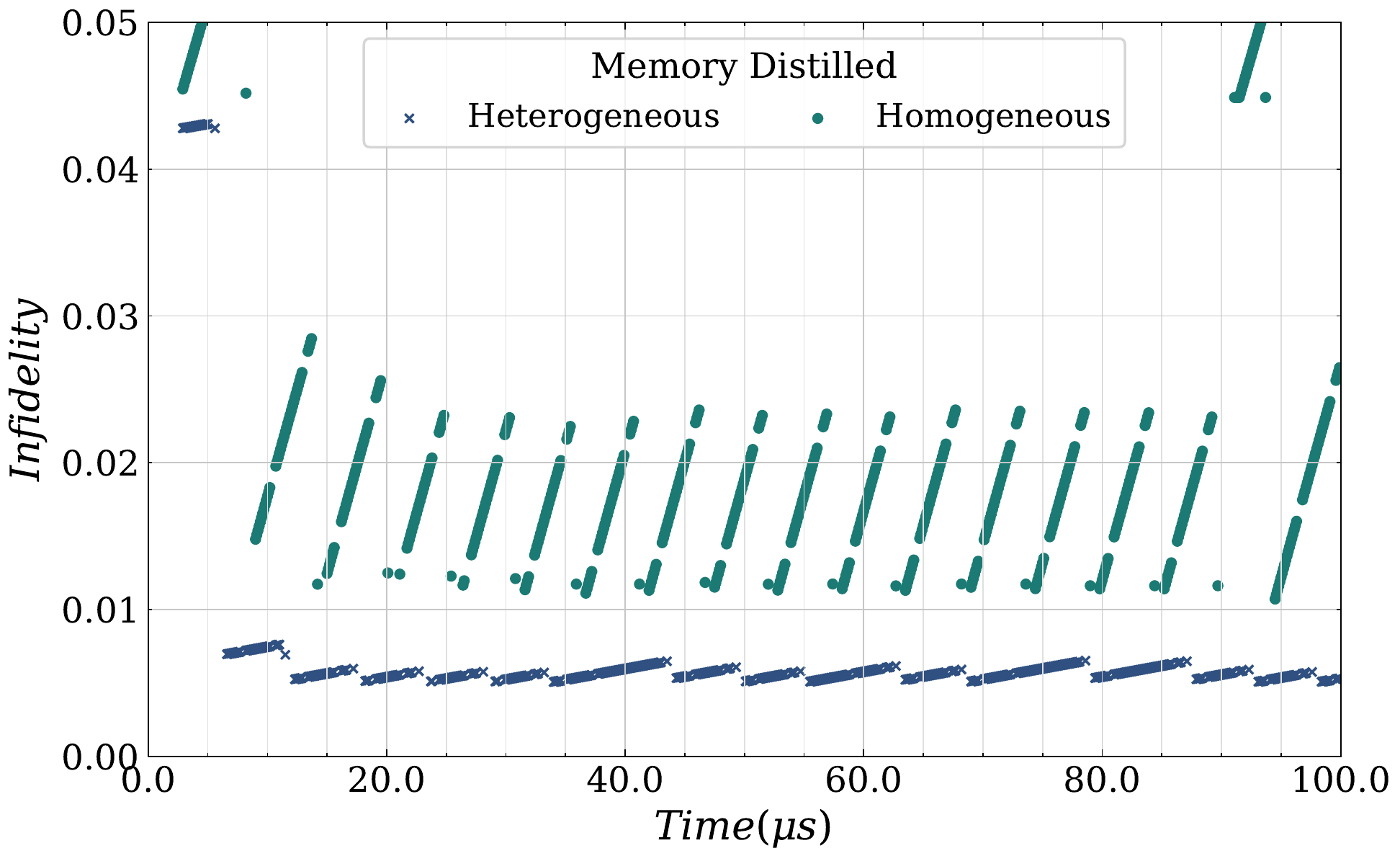}
    \caption{Distillation performance over time using both heterogeneous (blue) and homogeneous (green) systems, with probabilistic EP generation.}
    \label{fig:distill_action}
\end{figure}

\begin{figure}[]
    \centering
    \includegraphics[width=\columnwidth]{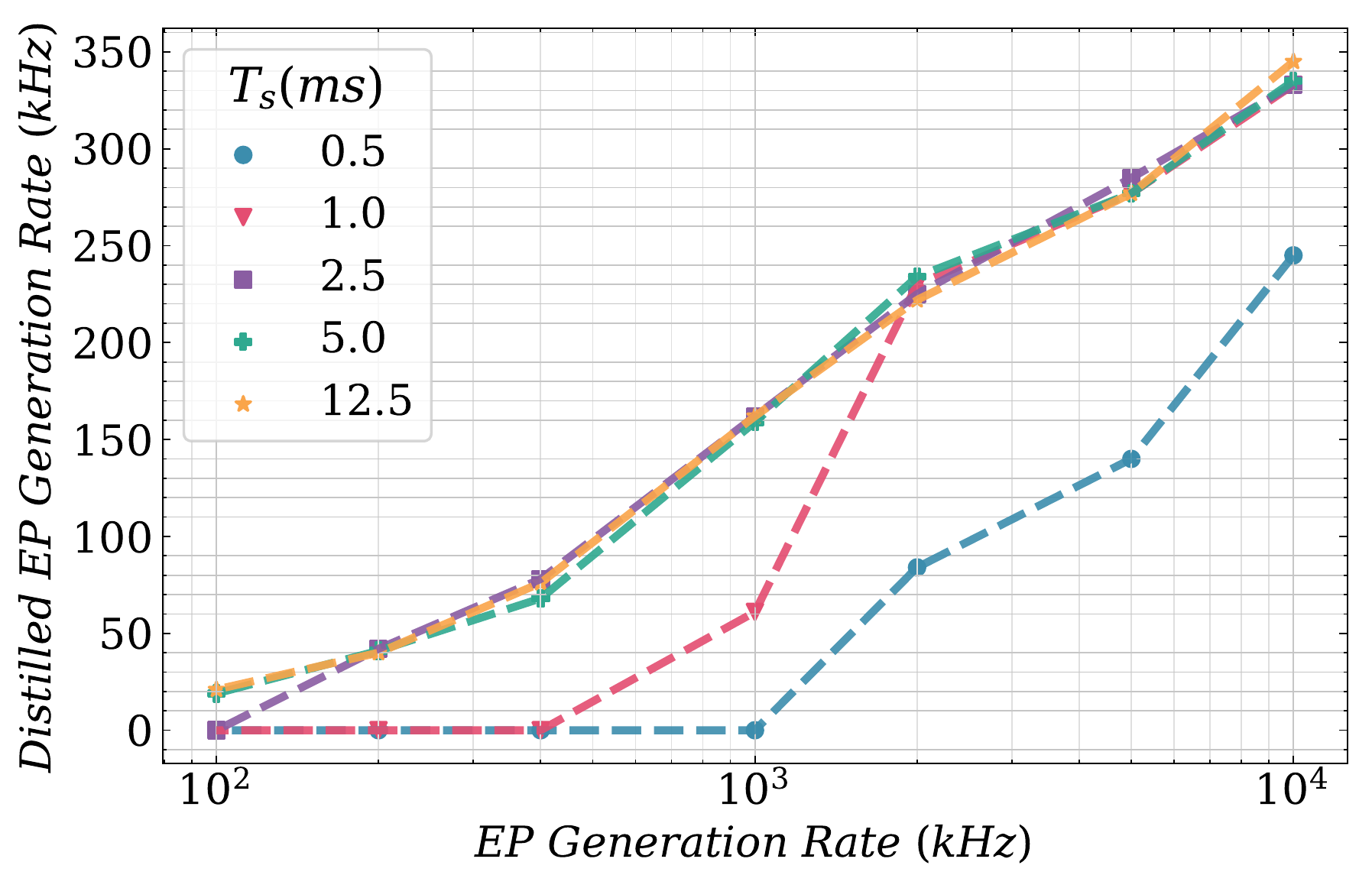} 
    \caption{EP distillation to fidelity $>.995$  rate for heterogeneous and homogeneous architectures as a function of EPR generation rate and storage coherence time $T_S$.}\label{fig:Distillation_Comparison}
\end{figure}

Simulation results for $T_s = 12.5$ms are shown in Fig.~\ref{fig:distill_action}. When compared to a homogeneous architecture, using both input and output memories allows the heterogeneous system to achieve a lower infidelity when distilling, as shown by the lower infidelity minima, as well as the ability to preserve that infidelity for longer, as shown by the shallower fidelity decay per unit time. Fig. \ref{fig:Distillation_Comparison} shows the rate of production of distilled EPs with fidelity above $0.995$ as a function of the generation rate of raw EPs and $T_S$. Distilled EP rates increase with $\lambda$ both because more pairs are available to distill in a given time and because of the reduced idle errors, which are even further reduced in a heterogeneous system. Heterogeneous systems with $T_S = 2.5$ms and greater outperform the homogeneous case with $T_S = T_C = 500\mu s$ by a factor of $2\times$ or more. 

The storage device performance required varies with the generation rate. For generation rates in excess of 10kHz, storage lifetimes of $T_S=1$ms are sufficient to achieve near-maximum performance, with diminishing returns for high lifetimes. For lower generation rates $T_S=2.5$ms or higher is required. Notably, for generation rates below 1000kHz, the homogeneous system composed of only compute devices fails to distill any pairs to threshold fidelity due to idling errors, while heterogeneous systems still allow generation on the order of 100kHz.





%% file: text/memory.tex
\subsection{Error Corrected Quantum Memory}\label{subsec:error-correction}

In a fault-tolerant quantum computer, error correction will be constantly running on noisy physical qubits, making it one of the most important subroutines~\cite{Martinis2015}. 
Programmatically, error correction requires two kinds of qubits, data and ancilla, with differing needs. Data qubits collectively store the state of the logical qubit, and so require higher coherence times, while ancilla qubits are used to perform error correction checks and hence need fast gates and readout.  

In this section, we consider two heterogeneous error correction architectures. The first explores planar surface code architectures, leveraging heterogeneity in coherence times for data and ancilla qubits. The second stores data qubits in high-capacity storage devices, leveraging both their long coherence times and many-to-one topology to provide a universal error correction architecture capable of implementing many QEC codes efficiently in a single device. 

QEC codes aim to perform `below threshold'~\cite{fowler2012surface}, wherein larger code distances correlate with a lower logical error rate or, for codes with only one distance, below `pseudo-threshold'~\cite{2017QS&T....2c5008C}, when the error rate of the system is below that of the hardware. As two-qubit gate errors are a major limitation of experimental architectures, in this subsection two-qubit gates are taken to have an error rate of $1\%$, which scaled-up experiments have reported~\cite{Acharya2023}. 


\subsubsection{Planar Surface Code}

\begin{figure}
    \centering
    \includegraphics[width = .9\columnwidth]{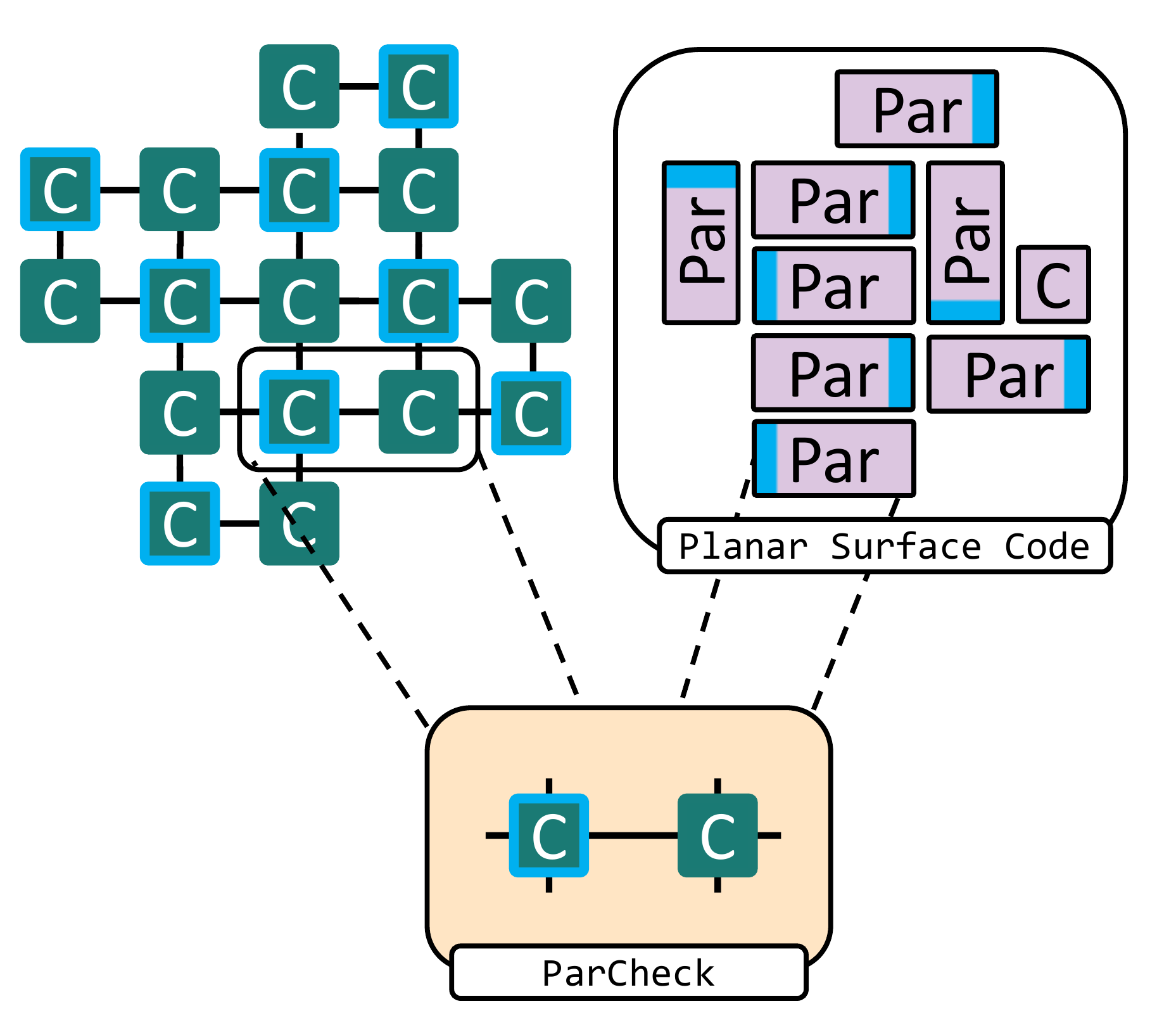}
    \caption{Hierarchical structure of a planar surface code quantum memory.}
    \label{fig:surface_code_hierarchy}
\end{figure}

Planar surface code error architectures have begun to be experimentally demonstrated and scaled in homogeneous systems  \cite{fowler2012surface, Krinner2022, Acharya2023}, with several heterogeneous approaches being proposed as well~\cite{Noh2022, Teoh2022, Hann2023MM}. 
In HetArch, the fundamental standard cell of the surface code is a data and ancilla qubit pair (Fig. \ref{fig:surface_code_hierarchy}). Since both undergo two-qubit gates too frequently to be storage devices, we can amend our approach and consider two classes of compute devices. The first class, optimized for high coherence, without readout, will store the data qubits; the second will be optimally coupled to a readout resonator for fast measurements and function as the ancilla. This directly maps to the parity check cell \texttt{ParCheck}. Because both are compute devices, we denote their coherence times (with $T_1 = T_2$) as $T_{CD}, T_{CA}$ for data and ancilla qubits, respectively. Note that due to the layout of the surface code, there will always be one more data than ancilla qubit. 

Simulation of a heterogeneous surface code is implemented using the Stim package \cite{StimRepo} which supports Monte Carlo experiments with a circuit-level noise model and allows for the performance of relatively large surface codes to be extrapolated from the properties of a standard cell: $T_{CA}, T_{CD}$, the single- and two-qubit gate durations and fidelities, and durations and fidelities of measurement and reset. These properties may be set to target the behavior of a future system, or, as in this case, extrapolate from current experimental performance. For this planar surface code example only, we take the baseline compute coherence to be 100$\mu$s (rather than 500$\mu$s as in Table \ref{tab:DeviceTable1}). This is to make the example more relevant, as the planar surface code is the only one of our examples that has already been implemented in a full experiment \cite{Krinner2022, Acharya2023}. 

\begin{figure}
    \centering
    \includegraphics[width=.9\columnwidth]{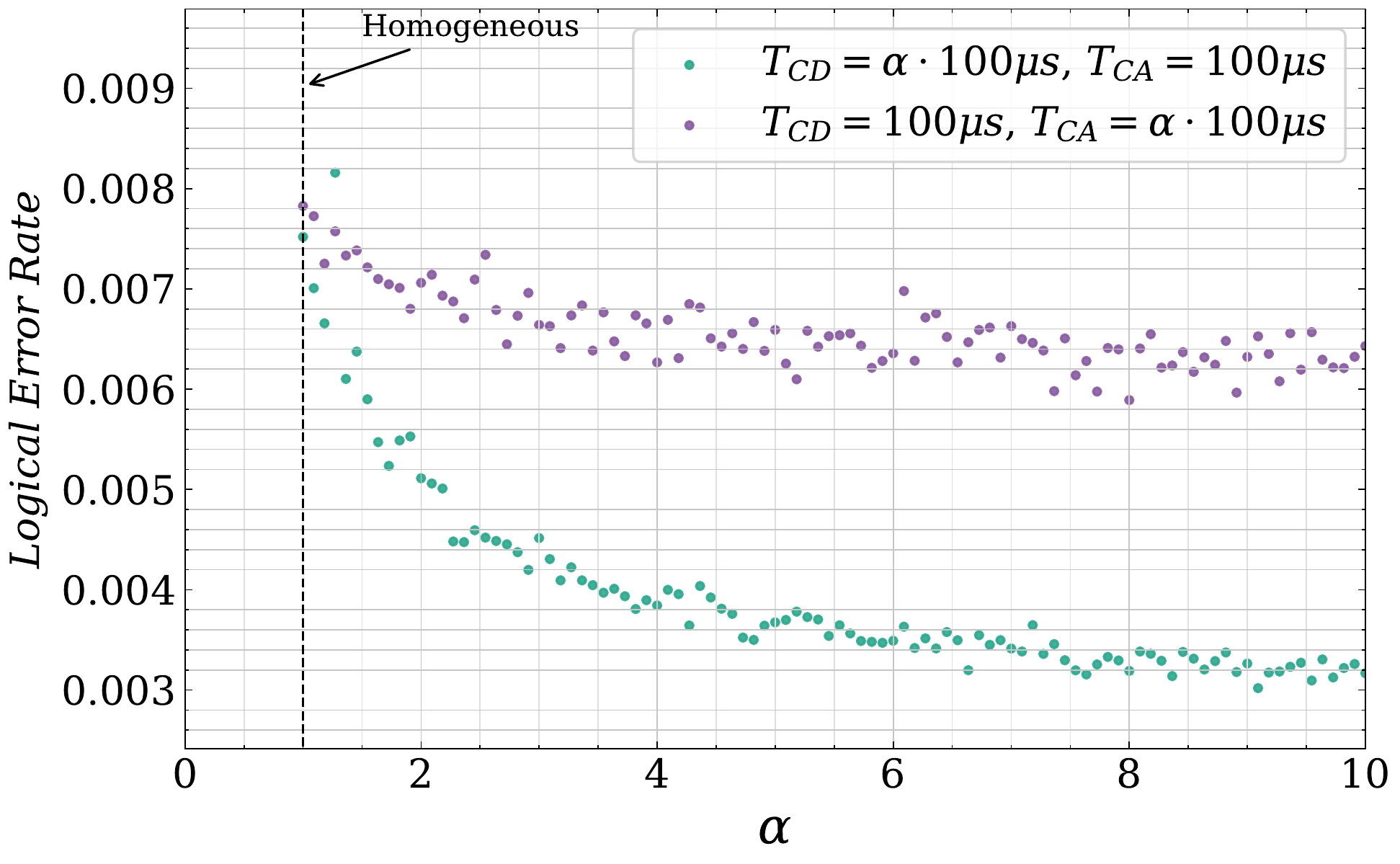}
    \caption{Logical error rate per cycle as a function of data and ancilla qubit coherence for distance $d$=13 surface code. In the homogeneous case, data qubit coherence $T_{CD}$ equals the ancilla qubit coherence $T_{CA}$. Increasing $T_{CD}$ by a factor of $\alpha$ leads to a greater improvement than increasing $T_{CA}$ by a factor of $\alpha$.}
\label{fig:surface_code_ancilla_vs_data}
\end{figure}

The results of increasing the data and ancilla qubit coherence times are shown in Fig. \ref{fig:surface_code_ancilla_vs_data}. As expected, increasing the coherence times of ancilla qubits does not reduce the error rate as much as increasing the coherence times of data qubits, largely because a major source of error is data qubit idling during the 1$\mu$s ancilla measurement. However, increasing the coherence times of data qubits leads to a 2.5x reduction in logical error rate for a data qubit coherence time of approximately 500$\mu$s, within experimental reach~\cite{Wang2022}, especially since data qubits do not require readout and can be designed with minimal leakage~\cite{Pechal2021}. Note that increasing data coherence times further leads to diminishing performance returns.  

\begin{figure}
    \centering
    \includegraphics[width=.9\columnwidth]{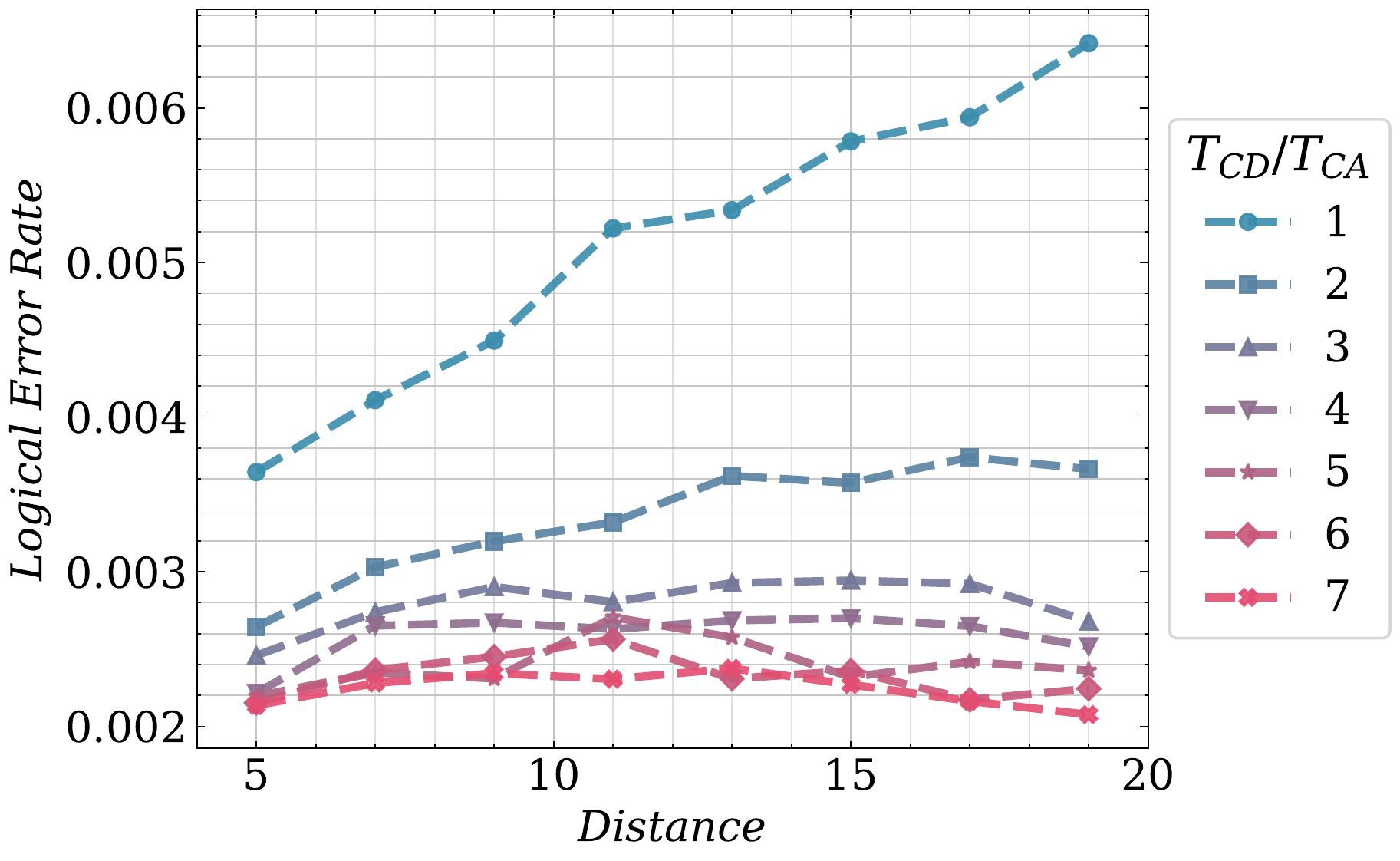}
    \caption{Logical error rate per cycle for code distances $d=5$ to distance $d=18$ as a function of the ratio $T_{CD}/T_{CA}$. When this ratio increases, the code moves below its threshold.}
\label{fig:SurfaceCodeResults}
\end{figure}

We also illustrate how the ratio $T_{CD}/T_{CA}$ affects the performance of the overall surface code. In Fig. \ref{fig:SurfaceCodeResults} we plot the logical error rate for code distances $d=5$ to distance $d=18$ as a function of the ratio $T_{CD}/T_{CA}$. When this ratio increases, the code moves below its threshold. Even with present-day two-qubit gate errors of 1\% and coherences on the order of 100$\mu$s, the surface code can benefit from a heterogeneous design where data qubits are optimized for coherence and ancilla qubits are optimized for readout~\cite{Acharya2023}.

\subsubsection{Universal Error Correction Module}\label{sec:UniversalErrorCorrection}
For homogeneous systems, the performance of quantum error correcting codes is highly dependent on matching the required topology of code checks to the connectivity of hardware, leading to severe restrictions on the variety of codes that can be implemented. Here we leverage the effective topology of high-capacity storage devices to create heterogeneous systems which are agnostic to the topology of a stabilizer QEC code, functioning as Universal Error Correction (UEC) modules at the cost of serializing QEC checks~\cite{Duckering2020}. 

\begin{figure}
    \centering
    \includegraphics[width = .8\columnwidth]{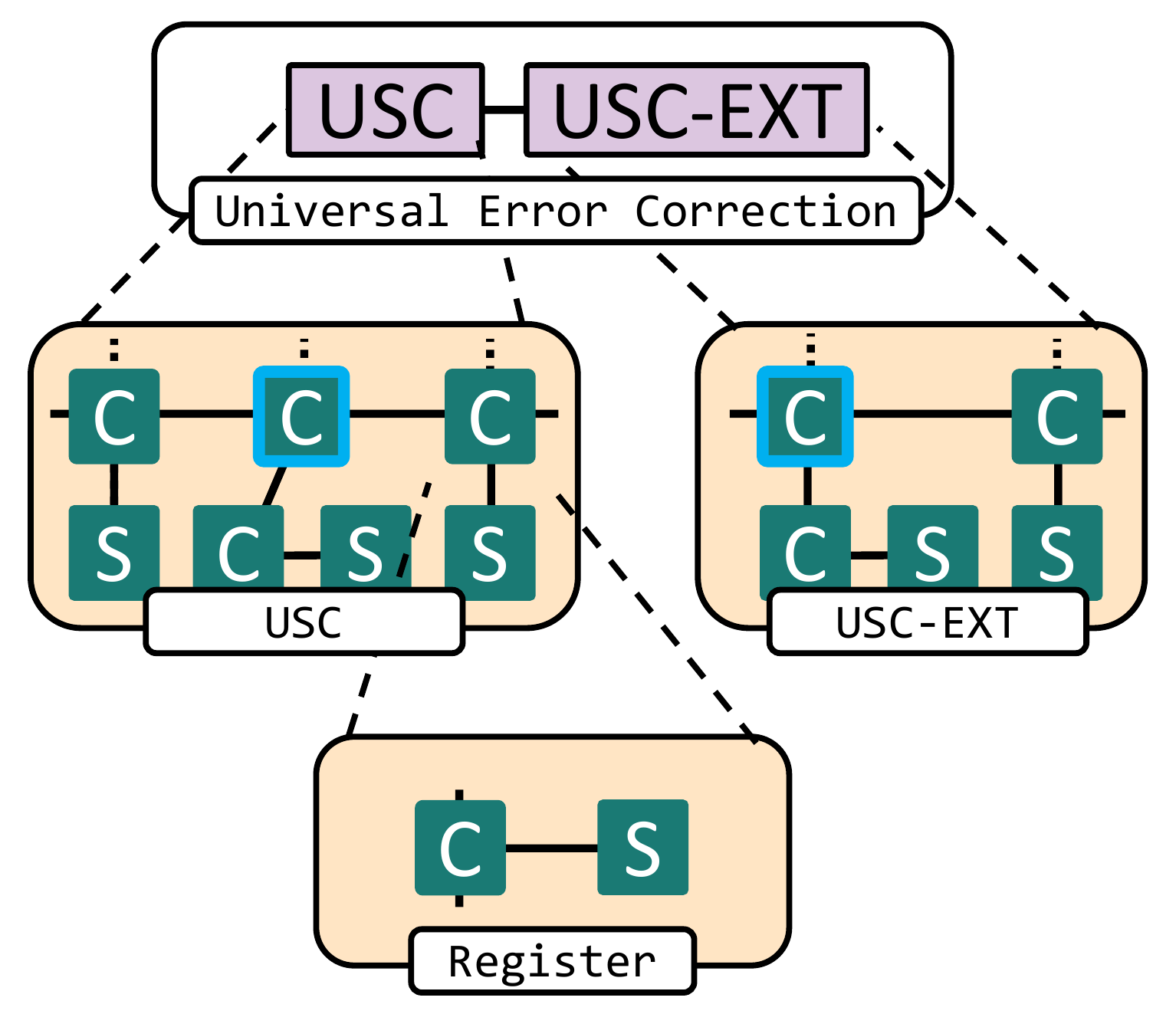}
    \caption{Hierarchical design of a Universal Error Correction module.}
    \label{fig:Universal-EC-Design}
\end{figure}

The basic function of the memory is to execute stabilizer checks. 
We store the data qubits in \texttt{Register} cells, keeping them in the multi-qubit storage devices when idling, and swapping them to compute elements only during stabilizer checks or logical gate operations. Because it contains many qubits, the storage device provides many-to-one coupling to compute device within the \texttt{Register} cell and also provides high-coherence while idling. Flag circuits may be used to ensure fault-tolerance~\cite{PhysRevX.10.011022, 2018Quant...2...53C}. 

The central element of a general error correction module, called the Universal Stabilizer Cell (\texttt{USC}), was originally described in Table~\ref{tab:StandardCellTable} and we now describe further. 
The design features three \texttt{Register} sub-cells (satisfying DR2) arranged around a central compute device with readout which will function as the ancilla qubit. This design represents the edge of the design space; \texttt{USC} designs with up to four \texttt{Register} cells were considered, and would allow even larger codes, but would exhaust the connectivity of the central ancilla compute (DR1), which is needed to connect to other cells or modules. Furthermore, this architecture allows for minimal number of internal connections (DR3) and readout qubits (DR4). Three additional outgoing connections are available if greater connectivity is required. As shown in Fig. \ref{fig:Universal-EC-Design}, defining an extension cell with only two \texttt{Register}s, \texttt{USC}-\texttt{ext}, allows multiple units to be chained together while still respecting DR1-4. Characterizing the \texttt{USC} as a subcell, rather than a submodule, reflects the fact that detailed density matrix simulations will need to be carried out spanning a \texttt{USC} and one or more \texttt{USC}-\texttt{EXT}s. 

The UEC module is able to adapt to any stabilizer code up to 30 qubits, and, with \text{USC}-\text{Ext}s added, to any code that can be partitioned in 1D for larger sizes. What enables this is the serialized error correction checks, in effect trading QEC code flexibility for long execution time and so requires long-lived storage~\cite{Gottesman:2022jlv}. Qubit assignment to the various \texttt{Register} devices, as well as the gate schedule, can be highly optimized. This first study utilizes a brute-force search to optimize assignments, limiting the model to 30 data qubits, and  then outputs a schedule of operations which seeks maximum possible parallelism while minimizing time each data qubit spends outside storage devices.
Replacing the brute-force search with scalable approaches may be the subject of future work.

\begin{figure}
    \centering
    \includegraphics[width = .9\columnwidth]{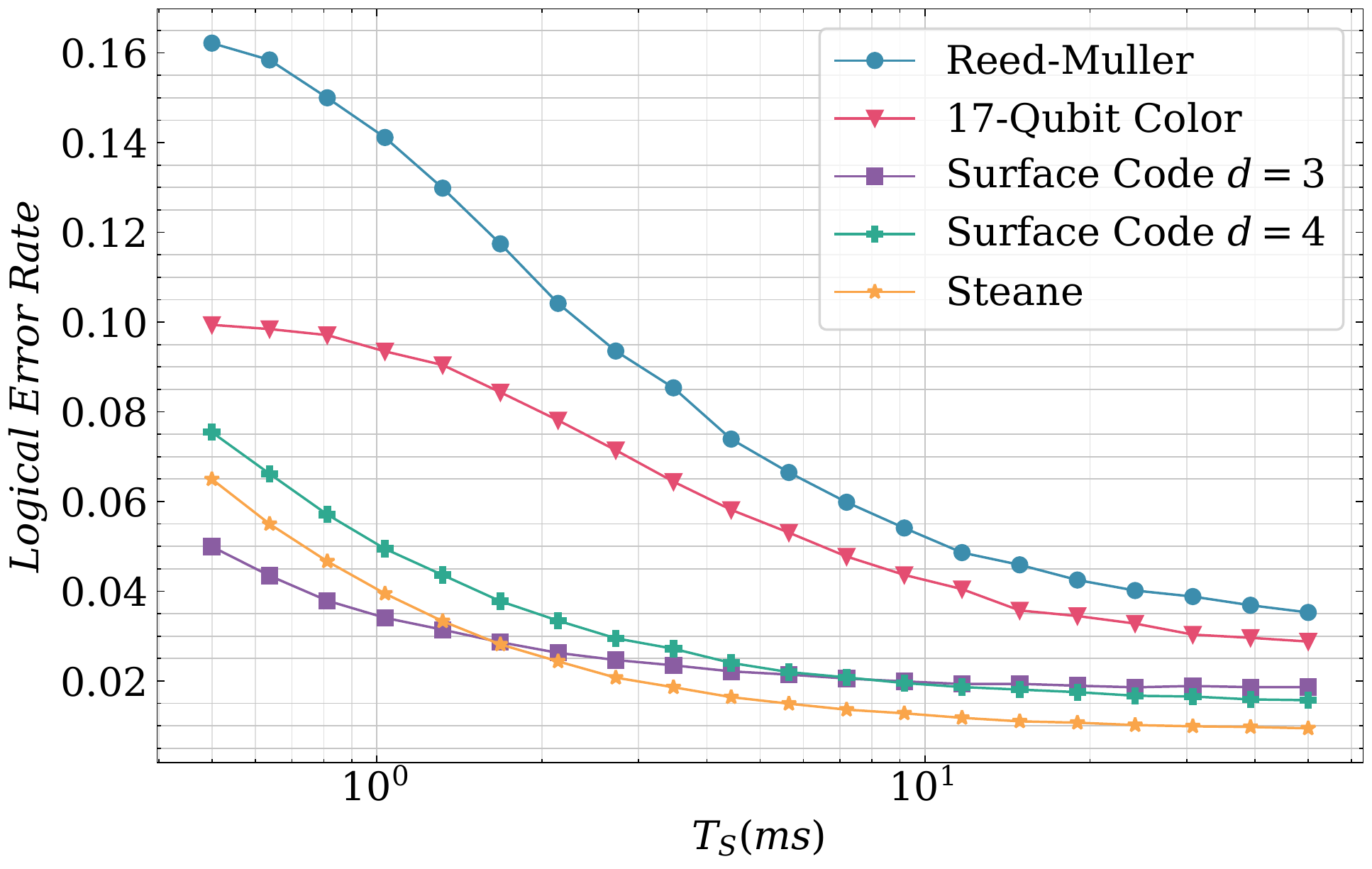}
    \caption{Performance of selected QEC codes on the Universal Error Correction Module.}
    \label{fig:GeneralErrorCorrection}
\end{figure}

\begin{table}[]
    \centering
    \begin{tabular}{|c|c|c|c|c|}\hline 
        Code & PT & Het. & Hom. & Red.\\ \hline\hline
        RM & 0.0254 & 0.0353 &  0.1660 & \cellcolor{lightgreen}4.7x\\ \hline
        17QCC & 0.1608 & 0.0284 & 0.0990 & \cellcolor{lightgreen}3.5x \\ \hline
        ST & 0.1291 & 0.0097 & 0.1034 & \cellcolor{lightgreen}10.7x \\\hline
        SC3 & - & 0.0186 & 0.0061 & \cellcolor{lightred}.3x \\\hline
        SC4 & - & 0.0158 & 0.0092 & \cellcolor{lightred}.6x\\\hline
    \end{tabular}
    \caption{QEC code, pseudothreshold (PT), logical error rates, and error reduction (Red.) for the heterogeneous (Het.)  architecture with $T_S = 50$ms and homogeneous (Hom.) architecture.}
    \label{tab:UEC_results}
\end{table}

Fig. \ref{fig:GeneralErrorCorrection} shows the performance of several QEC codes on the Universal Error Correction module simulated using STIM~\cite{StimRepo}. These codes are the surface code with $d=3, 4$ (SC3, SC4), the Steane 7-qubit code (ST), the 17-qubit Color code (17QCC), and the 15-qubit Reed-Muller (RM) code~\cite{bravyi2015doubled}. The surface codes are designed for a square lattice, and so fit naturally on the homogeneous architecture, while the other three are non-square (and the Reed-Muller is non-planar). For these small codes, two register units, and hence a single \texttt{USC} cell, are sufficient to maximize the parallelism of the checks while minimizing the number of \texttt{SWAP} operations that are needed. Performance of the codes, along with their pseudothresholds~\cite{2017QS&T....2c5008C} is listed in Table \ref{tab:UEC_results}. 
Thresholds for the surface code assume parallel stabilizer checks, and so do not apply to this serial execution, but we can see that the $d=4$ surface code outperforms the $d=3$ code, suggesting that the system is below threshold. The 17-qubit Color and Steane codes achieve error rates well below pseudo-threshold, while the Reed-Muller code misses pseudothreshold by approximately $50\%$.
Because the surface code is native to the square lattice, the UEC module underperforms the homogeneous architecture by 3.0x and 1.7x for $d=3$ and $d=4$, respectively. On the other hand, the Reed-Muller, 17-qubit Color, and Steane codes outperform the homogeneous architectures by 4.7x, 3.5x, and 10.7x, respectively.


%% file: text/teleportation.tex
\subsection{Code Teleportation}

\begin{figure}
    \centering
    \includegraphics[width=\columnwidth]{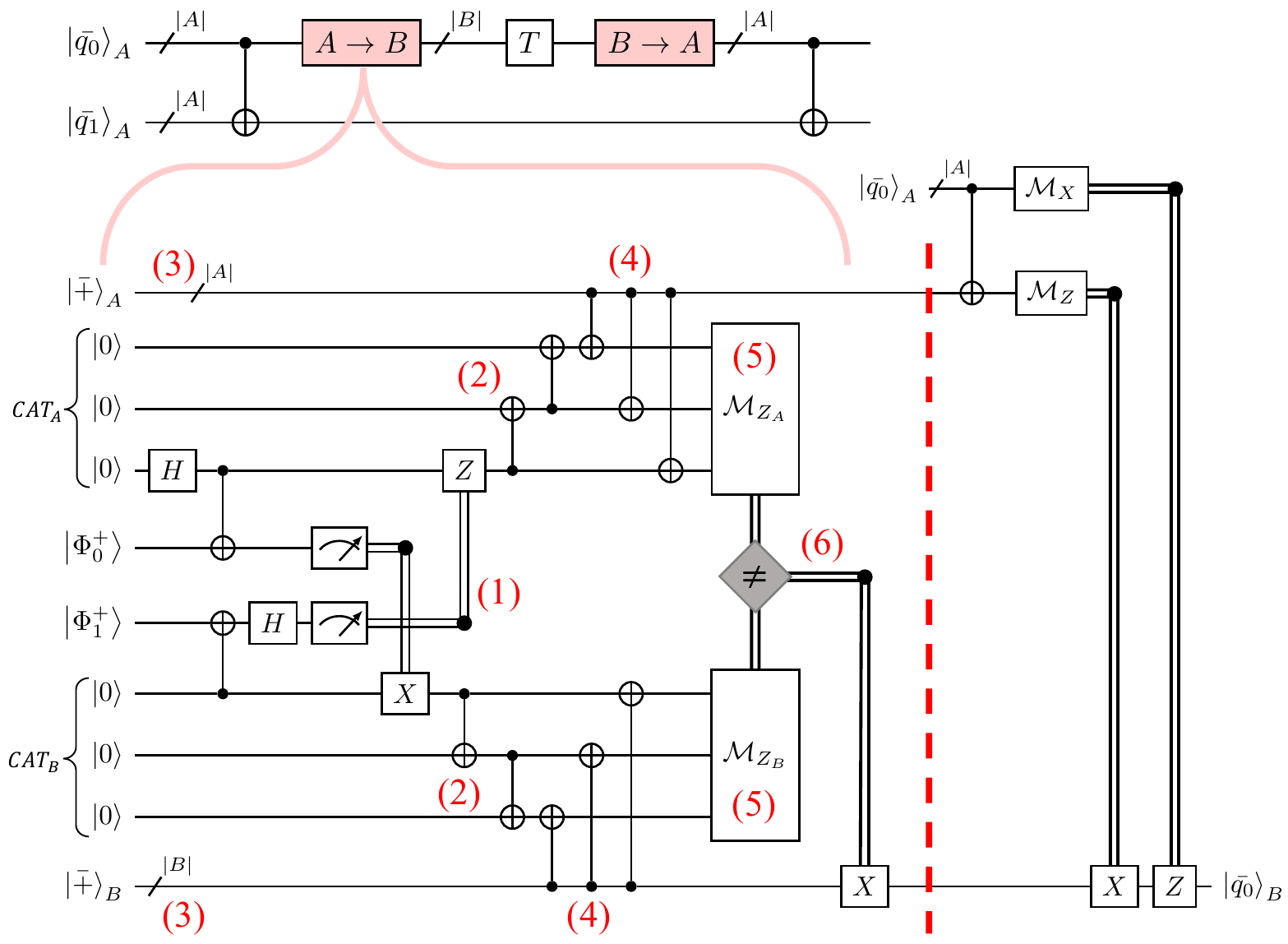}
    \caption{Example program (top) illustrating a compiler inserting code teleportation operations~\cite{choi2013cost} (bottom).
    }
    \label{fig:ct-compiler-example}
\end{figure}

\begin{figure}
    \centering
    \includegraphics[width=.9\columnwidth]{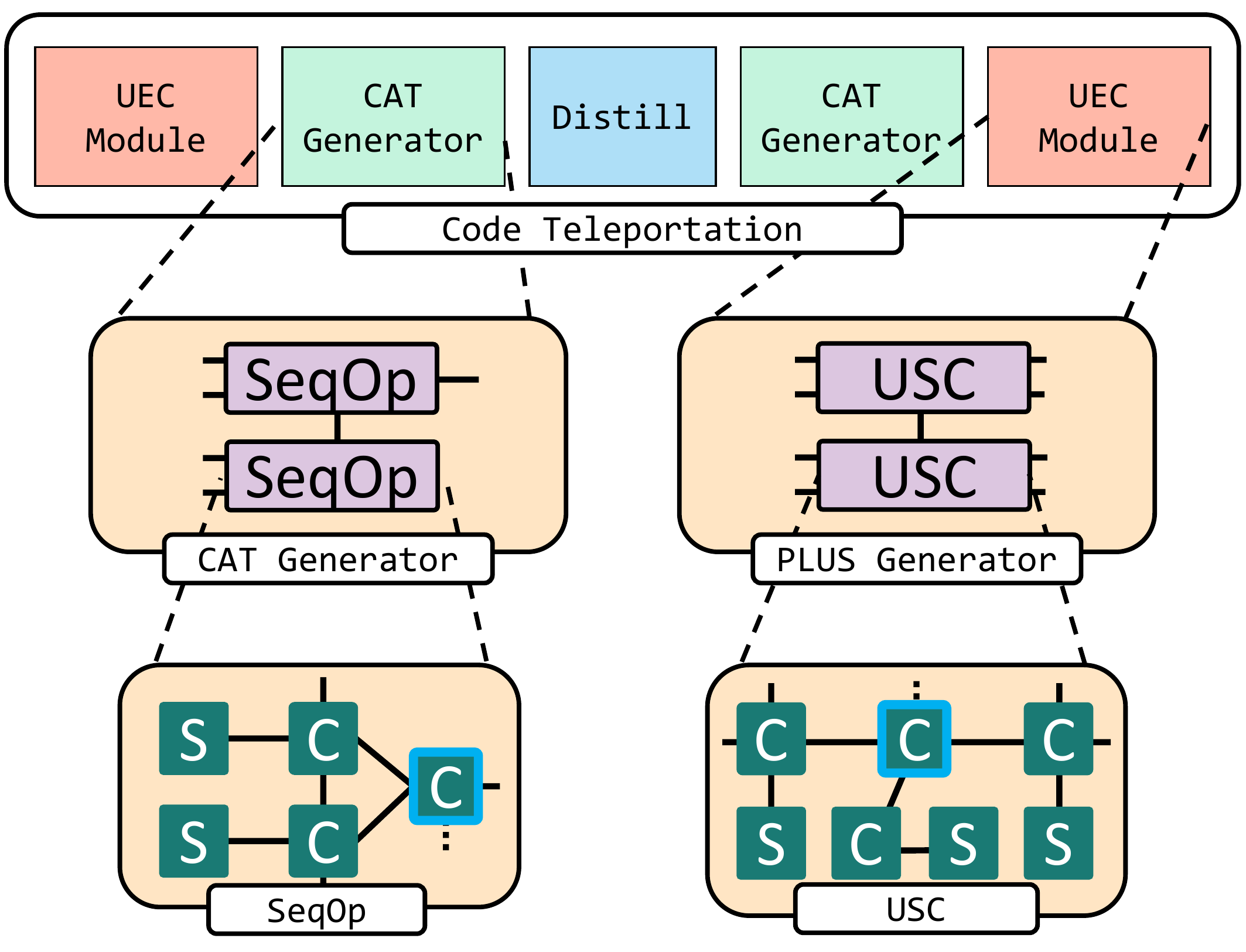}
    \caption{Hierarchical architecture of a Code Teleportation module.}
    \label{fig:code_teleportation_hierarchy}
\end{figure}

No single QEC code can provide a fault-tolerant gate set which is simultaneously universal and transversal \cite{eastin2009restrictions, zeng2011transversality}. 
Alongside state injection \cite{bravyi2012magic, litinski2019magic} and stabilizer sequences \cite{bravyi2015doubled}, code teleportation (CT) \cite{choi2013cost} addresses this issue, while also allowing conversion between different depth codes~\cite{thaker2006quantum}. CT functions by preparing CT resource states in the form of Bell states between logical codes, i.e. $\ket{\Phi}_{AB}^+=\frac{1}{\sqrt{2}}\ket{\bar 0_A\bar 0_B}+\ket{\bar 1_A\bar 1_B}$ for codes $A$ and $B$, so that performing logical teleportation both teleports the state and changes the QEC code.

However, a module for CT state preparation must be able to adapt to the different needs of at least two codes; this is particularly true if one of the codes is to have transversal \texttt{T} gates, which have high-weight, non-planar checks~\cite{bravyi2015doubled} not easily realized in traditional arrays. In that case, not only must code teleportation adapt to two codes itself, but bridge between two physical architectures, likely introducing a weak link with additional noise.

In this section, we design a dedicated CT module that functions between \textit{any} two arbitrary stabilizer codes up to 30 qubits. The CT module leverage both the distillation and Universal Error Correction (UEC) designs as submodules, along with a new submodule for CAT state generation. In particular, the flexibility of the UEC module allows the resulting CT module to act between any two codes within a single physical architecture, while the distillation module bridges the weak link between the two sides of the computer. 

The six steps of CT are shown in Figure \ref{fig:ct-compiler-example}, following~\cite{choi2013cost}. The key resources to create a CT state are logical $\ket{\bar +}$ states in the $A$ and $B$ codes, a shared CAT state of size $|A|+|B|$, and EPs that will be used to entangle and verify the CAT state. First, EPs are created \circled{1} and used for remote gates~\cite{szkopek2006threshold} to create a full CAT state of size $|A| + |B|$ \circled{2}. Then, logical $\ket{\bar + }$ states are created \circled{3}, and parallel \texttt{CNOT} gates entangle the logical $\ket{\bar +}$ states with the CAT state \circled{4}. Finally a logical measurement is performed \circled{5} and correction applied if needed \circled{6}. At the end of this process (dashed red line in Figure \ref{fig:ct-compiler-example}~) the CT module will have successfully prepared the code teleportation state $\Phi^+_{AB}$ which can be consumed to teleport and switch the QEC of an input state.  

For creating a physical design, the creation of these resource states each constitutes a subroutine, and each receives hardware module. As shown in Fig. \ref{fig:code_teleportation_hierarchy}, there are a total of five distinct submodules: a Distillation module, two CAT generators, and two UEC module, with the Distillation and UEC modules described in Sections \ref{sec:entanglement-distillation} and \ref{sec:UniversalErrorCorrection}. 

The new component of the CT module is the CAT generator submodule. This subroutine requires many sequential \texttt{CNOT}s with the result verified by parity checks. The Sequential Operations Cell (\texttt{SeqOp}), shown in Fig.~\ref{fig:code_teleportation_hierarchy}, is optimized for this purpose, with two \texttt{Register} subcells connected by a triangle of compute units, one with readout. This \texttt{SeqOp} design was selected after many design cycles as it balances between sequential operation performance and connectivity offered to other cells and modules. Relative to the \texttt{USC}, it offers direct two-qubit gates between qubits stored in the \texttt{Register} cells, while still allowing parity checks. 

A major source of design complexity within the Code Teleportation module is the need to support long range interactions between many of the sub-components, but this is mitigated in this architecture by the built-in storage capabilities of the distillation, UEC, and CAT generation sub-modules. At the same time, the universal capabilities of the UEC module allow the CT module to teleport between any two codes, regardless of their underlying topology, with an output from the ancilla compute of the UEC module allowing for transversal logical gates with an external module. The price of this flexibility is the serialization of stabilizer checks, which can only be afforded with long storage coherence times. 

\begin{figure}
    \centering
    \includegraphics[width=.9\columnwidth]{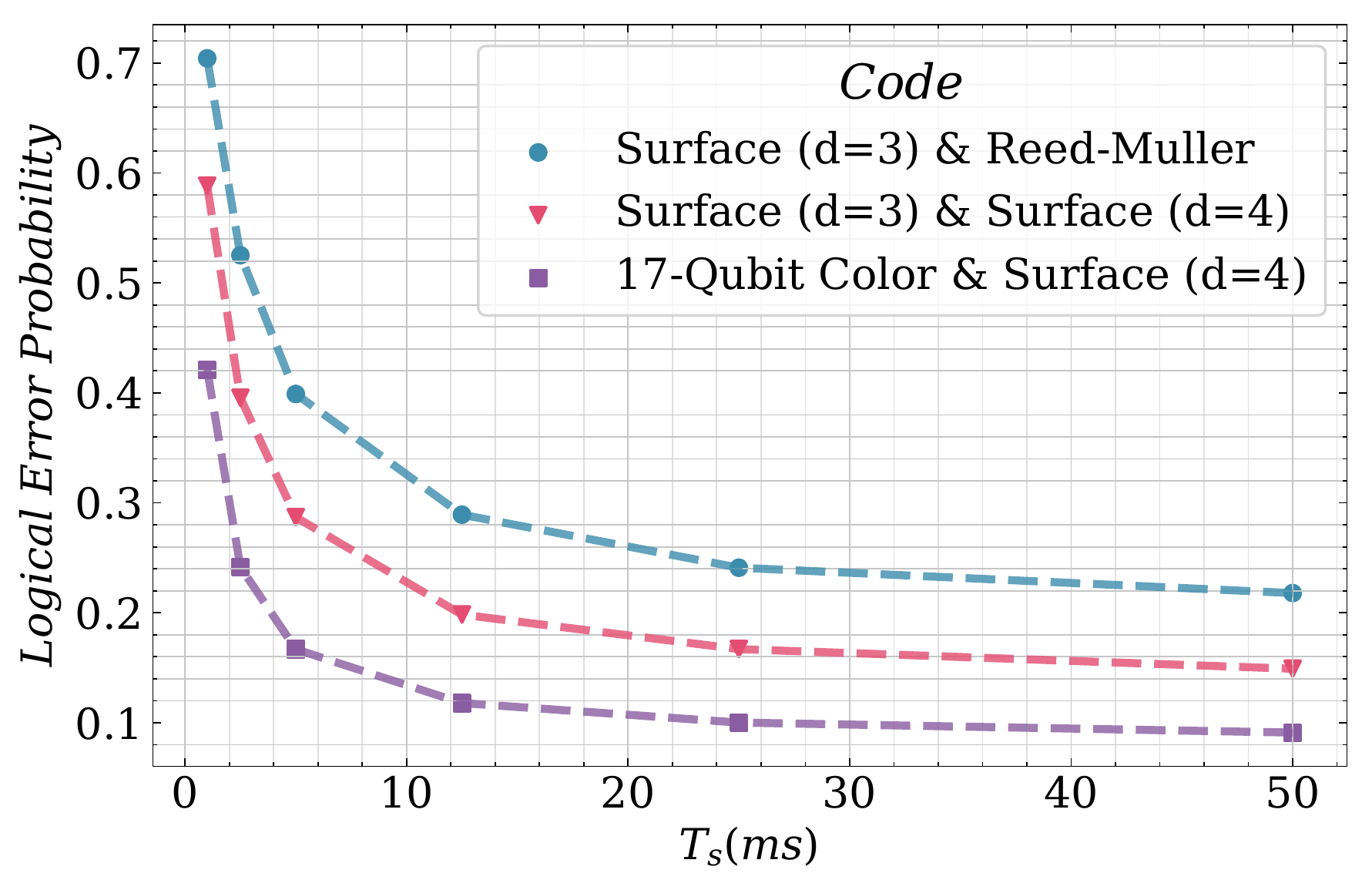}
    \caption{Code Teleportation performance using a heterogenous architecture between two codes. EP generation rate is 1000kHz, with a target distillation fidelity of 99.5\%. Homogeneous results are presented in Table \ref{tab:error_probabilities_ct}.}
    \label{fig:code_teleportation_performance}
\end{figure}

To evaluate the CT module, a target fidelity of 0.995 is set for the Distillation sub-module. Simulation then proceeds as in Section \ref{sec:entanglement-distillation} for the Distillation submodule and Section \ref{sec:UniversalErrorCorrection} for the UEC submodule. To simulate CAT state generation with realistic noise, a modular approach is also taken, modeling one code at a time, abstracting the remote \texttt{CNOT} gate~\cite{Ang2022}, and breaking the modeling of a single CAT generation into smaller CAT states, with multiplicative compounding fidelities. The parallel \texttt{CNOT} gate from step \circled{4} between the CAT and $\ket{\bar+}$ states is treated by similarly modeling errors in the CAT generator and UEC modules. For the logical $\ket{\bar +}$ state, the same approach described in Section~\ref{subsec:error-correction} yields logical error probability. To create the module level error model, independent error rates~\cite{10.1063/1.1499754} are summed.

Figure \ref{fig:code_teleportation_performance} shows the logical error probability in the prepared CT state, where the error probability decreases significantly as storage lifetime increases. The logical error rate performance is best for the largest codes with low (pseudo-) thresholds, namely the $d=4$ surface code and 17-qubit color code. The demand for high lifespan EP pair distillation, idling errors from CAT state parity checks, and errors during $\ket{\bar +}$ state stabilizer measurements reveal the substantial benefits of long-lived storage for code teleportation. For the simpler surface and Steane codes, $T_s$ times above $10$ms lead to diminishing returns, while for the more complicated codes, $T_s$ times above $50$ms, beyond what we consider for near term devices, may be advantageous.

Furthermore, Table \ref{tab:error_probabilities_ct} shows that heterogeneous CT systems outperform their homogeneous counterparts for every pair of codes studied. The expected advantage for non-planar codes is due to the UEC module's higher efficiency. Surprisingly, even for planar codes, heterogeneous systems outperform homogeneous ones, thanks to the enhanced performance of the distillation and CAT generation modules. Notably, some homogeneous experiments were unable to achieve the 99.5\% fidelity EP pair target, as the EP distillation submodule failed to reach the desired fidelity with EP generation rates of 1000kHz. The most significant reduction occurs between Reed-Muller and Surface Code (d=3), yielding a 2.96x reduction in logical error probability from 0.500 (essentially mixed) to 0.169. On average, error probabilities are reduced by 2.33x, with a minimum reduction of 1.60x, showcasing the remarkable potential of heterogeneous architectures.

\begin{table}[]
\small
    \centering
    \begin{tabular}{|c|c|c|c|c|c|}
        \hline
        & RM & 17Q CC & ST & SC (d=3) & SC (d=4) \\ \hline\hline
        RM & - & \cellcolor{lightgreen}0.284 & \cellcolor{lightgreen}0.189 & \cellcolor{lightgreen}0.169 & \cellcolor{lightgreen}0.202 \\ \hline
        17Q CC & \cellcolor{lightred}0.700 & - & \cellcolor{lightgreen}0.221 & \cellcolor{lightgreen}0.202 & \cellcolor{lightgreen}0.234 \\ \hline
        ST & \cellcolor{lightred}0.548 & \cellcolor{lightred}0.516 & - & \cellcolor{lightgreen}0.096 & \cellcolor{lightgreen}0.132 \\ \hline
        SC (d=3) & \cellcolor{lightred}0.500 & \cellcolor{lightred}0.465 & \cellcolor{lightred}0.192 & - & \cellcolor{lightgreen}0.111 \\ \hline
        SC (d=4) & \cellcolor{lightred}0.540 & \cellcolor{lightred}0.507 & \cellcolor{lightred}0.256 & \cellcolor{lightred}0.178 & - \\ \hline
    \end{tabular}
    \caption{Logical error probabilities of Code Teleportation, for heterogeneous (top right) and homogeneous systems (bottom left). Codes abbreviated as in Section \ref{sec:UniversalErrorCorrection}.}
    \label{tab:error_probabilities_ct}
\end{table}

%% file: text/conclusion.tex
\section{Conclusion}\label{sec:conclusion}


In this paper, we presented a novel study of heterogeneous quantum microarchitectures. By developing a hierarchical approach for aligning software and hardware needs, elucidating design rules for assembling devices into standard cells, and exploring design spaces for three heterogeneous quantum applications, we designed highly efficient systems tailored to their applications, delivering error rate reductions for entanglement distillation, error correction, and code teleportation of up to $10.7\times$. 


As we look to the future, we anticipate that experimental progress will yield even greater hardware heterogeneity. In atomic and ion systems, the use of multiple species or spin states presents tradeoffs involving multi-qubit gates, global control, and communication bottlenecks. Hybrid systems, comprising atoms~\cite{Young:2022cyz, Bluvstein:2021jsq}, ions~\cite{2019ApPRv...6b1314B, 2021Natur.592..209P, 2020AVSQS...2a4101K, 2021Natur.592..209P}, and superconducting devices, may be challenging to achieve experimentally but would combine high-speed superconducting compute capabilities with the long lifetimes of ions and atoms~\cite{doi:10.1146/annurev-conmatphys-030212-184253, 2016QuIP...15.5385D, 2016NatCo...710352S, 2021npjQI...7..121N}, albeit with a memory-compute bottleneck~\cite{2012PhRvL.108m0504K}. Even within superconducting systems, progress in routers~\cite{2023arXiv230201252M, Zhou2021} opens new connectivity, while variability within superconducting devices \cite{Gao2021} in fact offers functionality more like p-cells~\cite{kang2014cmos} in classical systems.

Through these future developments, we envision that the hierarchical HetArch methodology, breaking algorithms into subroutines to be executed by modules, and subroutines into operations to be executed by standard cells, will endure. However, the resulting architectures, and grouping of devices, will likely be completely different. Future quantum systems may encompass many levels of memory~\cite{thaker2006quantum}, with communication devices~\cite{1540969}, caches, and buffers. Compute regions may contain specialized hardware for Fourier transforms, modular multiplication, and local registers. In the burgeoning field of networked quantum systems, dedicated designs for both distillation modules and repeaters are likely to play an integral role. Through integrated heterogeneous design of software and hardware, the community is poised to explore a quantum computing design space as vast as that of classical computing.

%% file: main.bbl
\begin{thebibliography}{100}

\bibitem{Preskill2021}
John Preskill.
\newblock Quantum computing 40 years later.
\newblock 2021.

\bibitem{1996quant.ph..5011S}
P.W. Shor.
\newblock Fault-tolerant quantum computation.
\newblock In {\em Proceedings of 37th Conference on Foundations of Computer
  Science}, pages 56--65, 1996.

\bibitem{Shor1997}
Peter~W. Shor.
\newblock Polynomial-time algorithms for prime factorization and discrete
  logarithms on a quantum computer.
\newblock {\em SIAM Journal on Computing}, 26, 1997.

\bibitem{Feynman1982}
Richard~P. Feynman.
\newblock Simulating physics with computers.
\newblock {\em International Journal of Theoretical Physics}, 21, 1982.

\bibitem{byrnes2006simulating}
Tim Byrnes and Yoshihisa Yamamoto.
\newblock Simulating lattice gauge theories on a quantum computer.
\newblock {\em Physical Review A}, 73(2):022328, 2006.

\bibitem{kassal2011simulating}
Ivan Kassal, James~D Whitfield, Alejandro Perdomo-Ortiz, Man-Hong Yung, and
  Al{\'a}n Aspuru-Guzik.
\newblock Simulating chemistry using quantum computers.
\newblock {\em Annual review of physical chemistry}, 62:185--207, 2011.

\bibitem{reiher2017elucidating}
Markus Reiher, Nathan Wiebe, Krysta~M Svore, Dave Wecker, and Matthias Troyer.
\newblock Elucidating reaction mechanisms on quantum computers.
\newblock {\em Proceedings of the national academy of sciences},
  114(29):7555--7560, 2017.

\bibitem{havlivcek2019supervised}
Vojt{\v{e}}ch Havl{\'\i}{\v{c}}ek, Antonio~D C{\'o}rcoles, Kristan Temme,
  Aram~W Harrow, Abhinav Kandala, Jerry~M Chow, and Jay~M Gambetta.
\newblock Supervised learning with quantum-enhanced feature spaces.
\newblock {\em Nature}, 567(7747):209--212, 2019.

\bibitem{huang2020predicting}
Hsin-Yuan Huang, Richard Kueng, and John Preskill.
\newblock Predicting many properties of a quantum system from very few
  measurements.
\newblock {\em Nature Physics}, 16(10):1050--1057, 2020.

\bibitem{Beverland2022}
Michael~E. Beverland, Prakash Murali, Matthias Troyer, Krysta~M. Svore, Torsten
  Hoefler, Vadym Kliuchnikov, Guang~Hao Low, Mathias Soeken, Aarthi Sundaram,
  and Alexander Vaschillo.
\newblock Assessing requirements to scale to practical quantum advantage.
\newblock 2022.

\bibitem{PhysRevA.54.1098}
A.~R. Calderbank and Peter~W. Shor.
\newblock Good quantum error-correcting codes exist.
\newblock {\em Phys. Rev. A}, 54:1098--1105, Aug 1996.

\bibitem{nielsen2010quantum}
M.A. Nielsen and I.L. Chuang.
\newblock {\em Quantum Computation and Quantum Information: 10th Anniversary
  Edition}.
\newblock Cambridge University Press, 2010.

\bibitem{2021RvMP...93b5005B}
Alexandre {Blais}, Arne~L. {Grimsmo}, S.~M. {Girvin}, and Andreas {Wallraff}.
\newblock {Circuit quantum electrodynamics}.
\newblock {\em Reviews of Modern Physics}, 93(2):025005, April 2021.

\bibitem{Place2021}
Alexander~PM Place, Lila~VH Rodgers, Pranav Mundada, Basil~M Smitham, Mattias
  Fitzpatrick, Zhaoqi Leng, Anjali Premkumar, Jacob Bryon, Andrei Vrajitoarea,
  Sara Sussman, et~al.
\newblock New material platform for superconducting transmon qubits with
  coherence times exceeding 0.3 milliseconds.
\newblock {\em Nature communications}, 12(1):1--6, 2021.

\bibitem{Wang2022}
Chenlu Wang, Xuegang Li, Huikai Xu, Zhiyuan Li, Junhua Wang, Zhen Yang, Zhenyu
  Mi, Xuehui Liang, Tang Su, Chuhong Yang, Guangyue Wang, Wenyan Wang, Yongchao
  Li, Mo~Chen, Chengyao Li, Kehuan Linghu, Jiaxiu Han, Yingshan Zhang, Yulong
  Feng, Yu~Song, Teng Ma, Jingning Zhang, Ruixia Wang, Peng Zhao, Weiyang Liu,
  Guangming Xue, Yirong Jin, and Haifeng Yu.
\newblock Towards practical quantum computers: transmon qubit with a lifetime
  approaching 0.5 milliseconds.
\newblock {\em npj Quantum Information}, 8:3, 2022.

\bibitem{arute2019quantum}
Frank Arute, Kunal Arya, Ryan Babbush, Dave Bacon, Joseph~C Bardin, Rami
  Barends, Rupak Biswas, Sergio Boixo, Fernando~GSL Brandao, David~A Buell,
  et~al.
\newblock Quantum supremacy using a programmable superconducting processor.
\newblock {\em Nature}, 574(7779):505--510, 2019.

\bibitem{wu2021application}
Sau~Lan Wu, Jay Chan, Wen Guan, Shaojun Sun, Alex Wang, Chen Zhou, Miron Livny,
  Federico Carminati, Alberto Di~Meglio, Andy~CY Li, et~al.
\newblock Application of quantum machine learning using the quantum variational
  classifier method to high energy physics analysis at the lhc on ibm quantum
  computer simulator and hardware with 10 qubits.
\newblock {\em Journal of Physics G: Nuclear and Particle Physics},
  48(12):125003, 2021.

\bibitem{Lei2020}
Chan~U Lei, Lev Krayzman, Suhas Ganjam, Luigi Frunzio, and Robert~J.
  Schoelkopf.
\newblock High coherence superconducting microwave cavities with indium bump
  bonding.
\newblock {\em Applied Physics Letters}, 116:154002, 4 2020.

\bibitem{Chakram2021}
Srivatsan Chakram, Andrew~E. Oriani, Ravi~K. Naik, Akash~V. Dixit, Kevin He,
  Ankur Agrawal, Hyeokshin Kwon, and David~I. Schuster.
\newblock Seamless high- q microwave cavities for multimode circuit quantum
  electrodynamics.
\newblock {\em Physical Review Letters}, 127, 2021.

\bibitem{Ganjam2023MM}
Suhas~S Ganjam, Yanhao Wang, Yao Lu, Archan Banerjee, Chan~U Lei, Lev Krayzman,
  Kim Kisslinger, Chenyu Zhou, Yichen Jia, Mingzhao Liu, Luigi Frunzio, and
  Robert~J Schoelkopf.
\newblock Characterizing microwave losses in superconducting quantum circuits:
  Part 2.
\newblock {\em Bulletin of the American Physical Society}, 2023.

\bibitem{2023arXiv230206442M}
Ofir {Milul}, Barkay {Guttel}, Uri {Goldblatt}, Sergey {Hazanov}, Lalit~M.
  {Joshi}, Daniel {Chausovsky}, Nitzan {Kahn}, Engin {{\c{C}}ifty{\"u}rek},
  Fabien {Lafont}, and Serge {Rosenblum}.
\newblock {A superconducting quantum memory with tens of milliseconds coherence
  time}.
\newblock {\em arXiv e-prints}, page arXiv:2302.06442, February 2023.

\bibitem{Martinis2015}
John~M. Martinis.
\newblock Qubit metrology for building a fault-tolerant quantum computer.
\newblock {\em npj Quantum Information}, 1, 2015.

\bibitem{thaker2006quantum}
Darshan~D Thaker, Tzvetan~S Metodi, Andrew~W Cross, Isaac~L Chuang, and
  Frederic~T Chong.
\newblock Quantum memory hierarchies: Efficient designs to match available
  parallelism in quantum computing.
\newblock {\em ACM SIGARCH Computer Architecture News}, 34(2):378--390, 2006.

\bibitem{metodi2006quantum}
T.S. Metodi and F.T. Chong.
\newblock {\em Quantum Computing for Computer Architects}.
\newblock Synthesis lectures in computer architecture. Morgan \& Claypool
  Publishers, 2006.

\bibitem{10.1145/2903150.2906827}
X.~Fu, L.~Riesebos, L.~Lao, C.~G. Almudever, F.~Sebastiano, R.~Versluis,
  E.~Charbon, and K.~Bertels.
\newblock A heterogeneous quantum computer architecture.
\newblock In {\em Proceedings of the ACM International Conference on Computing
  Frontiers}, CF '16, page 323–330, New York, NY, USA, 2016. Association for
  Computing Machinery.

\bibitem{2019Quant...3..205L}
Daniel {Litinski}.
\newblock {Magic State Distillation: Not as Costly as You Think}.
\newblock {\em Quantum}, 3:205, December 2019.

\bibitem{2019arXiv190509749G}
Craig {Gidney} and Martin {Eker{\r{a}}}.
\newblock {How to factor 2048 bit RSA integers in 8 hours using 20 million
  noisy qubits}.
\newblock {\em arXiv e-prints}, page arXiv:1905.09749, May 2019.

\bibitem{2020arXiv201100028S}
Yunong {Shi}, Pranav {Gokhale}, Prakash {Murali}, Jonathan~M. {Baker}, Casey
  {Duckering}, Yongshan {Ding}, Natalie~C. {Brown}, Christopher {Chamberland},
  Ali {Javadi Abhari}, Andrew~W. {Cross}, David~I. {Schuster}, Kenneth~R.
  {Brown}, Margaret {Martonosi}, and Frederic~T. {Chong}.
\newblock {Resource-Efficient Quantum Computing by Breaking Abstractions}.
\newblock {\em arXiv e-prints}, page arXiv:2011.00028, October 2020.

\bibitem{biasedNoise}
Jahan Claes, J.~Eli Bourassa, and Shruti Puri.
\newblock Tailored cluster states with high threshold under biased noise.
\newblock {\em npj Quantum Information}, 9(1):9, 2023.

\bibitem{2022arXiv220813380F}
Sophia~Fuhui Lin, Sara Sussman, Casey Duckering, Pranav~S. Mundada, Jonathan~M.
  Baker, Rohan~S. Kumar, Andrew~A. Houck, and Frederic~T. Chong.
\newblock Let each quantum bit choose its basis gates.
\newblock {\em arXiv:2208.13380; In MICRO 2022: 55th IEEE/ACM International
  Symposium on Microarchitecture, 17 pages, 7 figures}, 2022.

\bibitem{2023arXiv230201252M}
Evan {McKinney}, Chao {Zhou}, Mingkang {Xia}, Michael {Hatridge}, and Alex~K.
  {Jones}.
\newblock {Parallel Driving for Fast Quantum Computing Under Speed Limits}.
\newblock {\em arXiv e-prints}, page arXiv:2302.01252, February 2023.

\bibitem{Hann2023MM}
Connor~T Hann, Christopher Chamberland, Harald Putterman, Joseph Iverson, Arne
  Grimsmo, Oskar Painter, Fernando Brandao, and Kyungjoo Noh.
\newblock Hybrid cat-transmon architecture for scalable, hardware-efficient
  quantum error correction.
\newblock {\em Bulletin of the American Physical Society}, 2023.

\bibitem{2020arXiv201204108C}
Christopher {Chamberland}, Kyungjoo {Noh}, Patricio {Arrangoiz-Arriola},
  Earl~T. {Campbell}, Connor~T. {Hann}, Joseph {Iverson}, Harald {Putterman},
  Thomas~C. {Bohdanowicz}, Steven~T. {Flammia}, Andrew {Keller}, Gil {Refael},
  John {Preskill}, Liang {Jiang}, Amir~H. {Safavi-Naeini}, Oskar {Painter}, and
  Fernando G.~S.~L. {Brand{\~a}o}.
\newblock {Building a fault-tolerant quantum computer using concatenated cat
  codes}.
\newblock {\em arXiv e-prints}, page arXiv:2012.04108, December 2020.

\bibitem{Chamberland2022}
Christopher Chamberland, Kyungjoo Noh, Patricio Arrangoiz-Arriola, Earl~T.
  Campbell, Connor~T. Hann, Joseph Iverson, Harald Putterman, Thomas~C.
  Bohdanowicz, Steven~T. Flammia, Andrew Keller, Gil Refael, John Preskill,
  Liang Jiang, Amir~H. Safavi-Naeini, Oskar Painter, and Fernando~G.S.L.
  Brandão.
\newblock Building a fault-tolerant quantum computer using concatenated cat
  codes.
\newblock {\em PRX Quantum}, 3, 2022.

\bibitem{2022arXiv221212077T}
James~D. {Teoh}, Patrick {Winkel}, Harshvardhan~K. {Babla}, Benjamin~J.
  {Chapman}, Jahan {Claes}, Stijn~J. {de Graaf}, John W.~O. {Garmon},
  William~D. {Kalfus}, Yao {Lu}, Aniket {Maiti}, Kaavya {Sahay}, Neel {Thakur},
  Takahiro {Tsunoda}, Sophia~H. {Xue}, Luigi {Frunzio}, Steven~M. {Girvin},
  Shruti {Puri}, and Robert~J. {Schoelkopf}.
\newblock {Dual-rail encoding with superconducting cavities}.
\newblock {\em arXiv e-prints}, page arXiv:2212.12077, December 2022.

\bibitem{2023Natur.616...50S}
V.~V. {Sivak}, A.~{Eickbusch}, B.~{Royer}, S.~{Singh}, I.~{Tsioutsios},
  S.~{Ganjam}, A.~{Miano}, B.~L. {Brock}, A.~Z. {Ding}, L.~{Frunzio}, S.~M.
  {Girvin}, R.~J. {Schoelkopf}, and M.~H. {Devoret}.
\newblock {Real-time quantum error correction beyond break-even}.
\newblock {\em \nat}, 616(7955):50--55, April 2023.

\bibitem{Duckering2020}
Casey Duckering, Jonathan~M. Baker, David~I. Schuster, and Frederic~T. Chong.
\newblock Virtualized logical qubits: A 2.5d architecture for error-corrected
  quantum computing.
\newblock volume 2020-October, 2020.

\bibitem{2022arXiv220504387M}
Evan {McKinney}, Mingkang {Xia}, Chao {Zhou}, Pinlei {Lu}, Michael {Hatridge},
  and Alex~K. {Jones}.
\newblock {Co-Designed Architectures for Modular Superconducting Quantum
  Computers}.
\newblock {\em arXiv e-prints}, page arXiv:2205.04387, May 2022.

\bibitem{2021PhRvL.127n0503G}
{\'E}lie {Gouzien} and Nicolas {Sangouard}.
\newblock {Factoring 2048-bit RSA Integers in 177 Days with 13 436 Qubits and a
  Multimode Memory}.
\newblock {\em Phys. Rev. Lett.}, 127(14):140503, October 2021.

\bibitem{PhysRevX.10.021060}
Christopher~S. Wang, Jacob~C. Curtis, Brian~J. Lester, Yaxing Zhang, Yvonne~Y.
  Gao, Jessica Freeze, Victor~S. Batista, Patrick~H. Vaccaro, Isaac~L. Chuang,
  Luigi Frunzio, Liang Jiang, S.~M. Girvin, and Robert~J. Schoelkopf.
\newblock Efficient multiphoton sampling of molecular vibronic spectra on a
  superconducting bosonic processor.
\newblock {\em Phys. Rev. X}, 10:021060, Jun 2020.

\bibitem{wangthesis}
Christopher Wang.
\newblock Bosonic quantum simulation in circuit quantum electrodynamics, 2022.

\bibitem{PhysRevResearch.3.043072}
Zohreh Davoudi, Norbert~M. Linke, and Guido Pagano.
\newblock Toward simulating quantum field theories with controlled phonon-ion
  dynamics: A hybrid analog-digital approach.
\newblock {\em Phys. Rev. Res.}, 3:043072, Oct 2021.

\bibitem{PhysRevX.10.011022}
Christopher Chamberland, Guanyu Zhu, Theodore~J. Yoder, Jared~B. Hertzberg, and
  Andrew~W. Cross.
\newblock Topological and subsystem codes on low-degree graphs with flag
  qubits.
\newblock {\em Phys. Rev. X}, 10:011022, Jan 2020.

\bibitem{Ang2022}
James Ang, Gabriella Carini, Yanzhu Chen, Isaac Chuang, Michael~Austin DeMarco,
  Sophia~E. Economou, Alec Eickbusch, Andrei Faraon, Kai-Mei Fu, Steven~M.
  Girvin, Michael Hatridge, Andrew Houck, Paul Hilaire, Kevin Krsulich, Ang Li,
  Chenxu Liu, Yuan Liu, Margaret Martonosi, David~C. McKay, James Misewich,
  Mark Ritter, Robert~J. Schoelkopf, Samuel~A. Stein, Sara Sussman, Hong~X.
  Tang, Wei Tang, Teague Tomesh, Norm~M. Tubman, Chen Wang, Nathan Wiebe,
  Yong-Xin Yao, Dillon~C. Yost, and Yiyu Zhou.
\newblock Architectures for multinode superconducting quantum computers.
\newblock 2022.

\bibitem{9355323}
Ang Li, Omer Subasi, Xiu Yang, and Sriram Krishnamoorthy.
\newblock Density matrix quantum circuit simulation via the bsp machine on
  modern gpu clusters.
\newblock In {\em SC20: International Conference for High Performance
  Computing, Networking, Storage and Analysis}, pages 1--15, 2020.

\bibitem{10.1063/1.1499754}
Eric Dennis, Alexei Kitaev, Andrew Landahl, and John Preskill.
\newblock {Topological quantum memory}.
\newblock {\em Journal of Mathematical Physics}, 43(9):4452--4505, 08 2002.

\bibitem{10.1145/3028687.3038873}
Mohamed Zahran.
\newblock Heterogeneous computing: Here to stay: Hardware and software
  perspectives.
\newblock {\em Queue}, 14(6):31–42, dec 2016.

\bibitem{tanenbaum1999structured}
A.S. Tanenbaum and J.R. Goodman.
\newblock {\em Structured Computer Organization}.
\newblock Prentice Hall international editions. Prentice Hall, 1999.

\bibitem{hall1992microprocessors}
D.V. Hall.
\newblock {\em Microprocessors and Interfacing: Programming and Hardware}.
\newblock McGraw-Hill computer science series. Glencoe, 1992.

\bibitem{1980aw...book.....M}
C.~{Mead} and L.~{Conway}.
\newblock {\em {Introduction to VLSI systems}}.
\newblock 1980.

\bibitem{weste1985principles}
Neil~HE Weste and Kamran Eshraghian.
\newblock {\em Principles of CMOS VLSI design: a systems perspective}.
\newblock Addison-Wesley Longman Publishing Co., Inc., 1985.

\bibitem{shankar2014vlsi}
R.~Shankar, E.B. Fernandez, and N.G. Einspruch.
\newblock {\em VLSI and Computer Architecture}.
\newblock ISSN. Elsevier Science, 2014.

\bibitem{IBMRoadmap}
IBM.
\newblock Ibm quantum roadmap.
\newblock Accessed: Sep. 20, 2022.

\bibitem{2014PhRvA..89b2317M}
C.~{Monroe}, R.~{Raussendorf}, A.~{Ruthven}, K.~R. {Brown}, P.~{Maunz}, L.~M.
  {Duan}, and J.~{Kim}.
\newblock {Large-scale modular quantum-computer architecture with atomic memory
  and photonic interconnects}.
\newblock {\em pra}, 89(2):022317, February 2014.

\bibitem{2023arXiv230108542T}
Andrew~K. {Tan}, Yuan {Liu}, Minh~C. {Tran}, and Isaac~L. {Chuang}.
\newblock {Error Correction of Quantum Algorithms: Arbitrarily Accurate
  Recovery Of Noisy Quantum Signal Processing}.
\newblock {\em arXiv e-prints}, page arXiv:2301.08542, January 2023.

\bibitem{1996quant.ph..8012K}
Emanuel {Knill} and Raymond {Laflamme}.
\newblock {Concatenated Quantum Codes}.
\newblock {\em arXiv e-prints}, pages quant--ph/9608012, August 1996.

\bibitem{10.1145/1330521.1330522}
Tzvetan~S. Metodi, Darshan~D. Thaker, Andrew~W. Cross, Isaac~L. Chuang, and
  Frederic~T. Chong.
\newblock High-level interconnect model for the quantum logic array
  architecture.
\newblock {\em J. Emerg. Technol. Comput. Syst.}, 4(1), apr 2008.

\bibitem{9167259}
Shesha Raghunathan and Leon Stok.
\newblock Eda and quantum computing: a symbiotic relationship?
\newblock {\em IEEE Design \& Test}, 37(6):71--78, 2020.

\bibitem{10.1145/3439706.3446900}
Leon Stok.
\newblock Eda and quantum computing: The key role of quantum circuits.
\newblock In {\em Proceedings of the 2021 International Symposium on Physical
  Design}, ISPD '21, page 111, New York, NY, USA, 2021. Association for
  Computing Machinery.

\bibitem{8587760}
Robert Wille, Austin Fowler, and Yehuda Naveh.
\newblock Computer-aided design for quantum computation.
\newblock In {\em 2018 IEEE/ACM International Conference on Computer-Aided
  Design (ICCAD)}, pages 1--6, 2018.

\bibitem{aspuru-guzik_cad}
Tim Menke, Florian H{\"a}se, Simon Gustavsson, Andrew~J. Kerman, William~D.
  Oliver, and Al{\'a}n Aspuru-Guzik.
\newblock Automated design of superconducting circuits and its application to
  4-local couplers.
\newblock {\em npj Quantum Information}, 7(1):49, 2021.

\bibitem{Kyaw:2020qcz}
Thi~Ha Kyaw, Tim Menke, Sukin Sim, Abhinav Anand, Nicolas P.~D. Sawaya,
  William~D. Oliver, Gian~Giacomo Guerreschi, and Al\'an Aspuru-Guzik.
\newblock {Quantum Computer-Aided Design: Digital Quantum Simulation of Quantum
  Processors}.
\newblock {\em Phys. Rev. Applied}, 16(4):044042, 2021.

\bibitem{8342103}
Yehuda Naveh, Elham Kashefi, James~R. Wootton, and Koen Bertels.
\newblock Theoretical and practical aspects of verification of quantum
  computers.
\newblock In {\em 2018 Design, Automation \& Test in Europe Conference \&
  Exhibition (DATE)}, pages 721--730, 2018.

\bibitem{10.1093/ietfec/e91-a.2.584}
Shiou-An Wang, Chin-Yung Lu, I-Ming Tsai, and Sy-Yen Kuo.
\newblock An xqdd-based verification method for quantum circuits.
\newblock {\em IEICE Trans. Fundam. Electron. Commun. Comput. Sci.},
  E91-A(2):584–594, feb 2008.

\bibitem{2021Quant...5..583G}
Peter {Groszkowski} and Jens {Koch}.
\newblock {Scqubits: a Python package for superconducting qubits}.
\newblock {\em Quantum}, 5:583, November 2021.

\bibitem{2022arXiv220608320P}
Sai {Pavan Chitta}, Tianpu {Zhao}, Ziwen {Huang}, Ian {Mondragon-Shem}, and
  Jens {Koch}.
\newblock {Computer-aided quantization and numerical analysis of
  superconducting circuits}.
\newblock {\em arXiv e-prints}, page arXiv:2206.08320, June 2022.

\bibitem{Qiskit_Metal}
Zlatko~K Minev, Thomas~G McConkey, Jeremy Drysdale, Priti Shah, Dennis Wang,
  Marco Facchini, Grace Harper, John Blair, Helena Zhang, Nick Lanzillo,
  Sagarika Mukesh, Will Shanks, Chris Warren, and Jay~M Gambetta.
\newblock {Qiskit Metal: An Open-Source Framework for Quantum Device Design
  {\&} Analysis}, 2021.

\bibitem{doi:10.1146/annurev-conmatphys-030212-184253}
Nikos Daniilidis and Hartmut H\"{a}ffner.
\newblock Quantum interfaces between atomic and solid-state systems.
\newblock {\em Annual Review of Condensed Matter Physics}, 4(1):83--112, 2013.

\bibitem{2016QuIP...15.5385D}
D.~{De Motte}, A.~R. {Grounds}, M.~{Reh{\'a}k}, A.~{Rodriguez Blanco},
  B.~{Lekitsch}, G.~S. {Giri}, P.~{Neilinger}, G.~{Oelsner}, E.~{Il'ichev},
  M.~{Grajcar}, and W.~K. {Hensinger}.
\newblock {Experimental system design for the integration of trapped-ion and
  superconducting qubit systems}.
\newblock {\em Quantum Information Processing}, 15(12):5385--5414, December
  2016.

\bibitem{2012PhRvL.108m0504K}
D.~{Kielpinski}, D.~{Kafri}, M.~J. {Woolley}, G.~J. {Milburn}, and J.~M.
  {Taylor}.
\newblock {Quantum Interface between an Electrical Circuit and a Single Atom}.
\newblock {\em \prl}, 108(13):130504, March 2012.

\bibitem{PhysRevLett.103.043603}
J.~Verd\'u, H.~Zoubi, Ch. Koller, J.~Majer, H.~Ritsch, and J.~Schmiedmayer.
\newblock Strong magnetic coupling of an ultracold gas to a superconducting
  waveguide cavity.
\newblock {\em Phys. Rev. Lett.}, 103:043603, Jul 2009.

\bibitem{2022arXiv220906841B}
Sergey {Bravyi}, Oliver {Dial}, Jay~M. {Gambetta}, Dario {Gil}, and Zaira
  {Nazario}.
\newblock {The Future of Quantum Computing with Superconducting Qubits}.
\newblock {\em arXiv e-prints}, page arXiv:2209.06841, September 2022.

\bibitem{2012arXiv1208.0391M}
C.~{Monroe}, R.~{Raussendorf}, A.~{Ruthven}, K.~R. {Brown}, P.~{Maunz}, L.~M.
  {Duan}, and J.~{Kim}.
\newblock {Large Scale Modular Quantum Computer Architecture with Atomic Memory
  and Photonic Interconnects}.
\newblock {\em arXiv e-prints}, page arXiv:1208.0391, August 2012.

\bibitem{2019NatCo..10.4692Z}
Kuan {Zhang}, Jayne {Thompson}, Xiang {Zhang}, Yangchao {Shen}, Yao {Lu},
  Shuaining {Zhang}, Jiajun {Ma}, Vlatko {Vedral}, Mile {Gu}, and Kihwan {Kim}.
\newblock {Modular quantum computation in a trapped ion system}.
\newblock {\em Nature Communications}, 10:4692, October 2019.

\bibitem{LaRacuente:2022xqq}
Nicholas LaRacuente, Kaitlin~N. Smith, Poolad Imany, Kevin~L. Silverman, and
  Frederic~T. Chong.
\newblock {Short-Range Microwave Networks to Scale Superconducting Quantum
  Computation}.
\newblock 1 2022.

\bibitem{9923784}
Kaitlin~N. Smith, Gokul~Subramanian Ravi, Jonathan~M. Baker, and Frederic~T.
  Chong.
\newblock Scaling superconducting quantum computers with chiplet architectures.
\newblock In {\em 2022 55th IEEE/ACM International Symposium on
  Microarchitecture (MICRO)}, pages 1092--1109, 2022.

\bibitem{6370030}
S.~Trimberger and R.~F.~W. Pease.
\newblock Reaching for the million-transistor chip.
\newblock {\em IEEE Spectrum}, 20(11):100--105, 1983.

\bibitem{alagic1978design}
S.~Alagic and M.A. Arbib.
\newblock {\em The Design of Well-Structured and Correct Programs}.
\newblock Monographs in Computer Science. Springer New York, 1978.

\bibitem{VIJEYAKUMAR20122186}
K.N. Vijeyakumar, V.~Sumathy, M.Gayathri Devi, S.~Tamilselvan, and Remya.R.
  Nair.
\newblock Design of hardware efficient high speed multiplier using modified
  ternary logic.
\newblock {\em Procedia Engineering}, 38:2186--2195, 2012.
\newblock INTERNATIONAL CONFERENCE ON MODELLING OPTIMIZATION AND COMPUTING.

\bibitem{QiskitMetal}
{Qiskit Metal}.
\newblock \url{https://qiskit.org/metal/ }, 2023.

\bibitem{2022arXiv220604990D}
Evan~E. {Dobbs}, Joseph~S. {Friedman}, and Alexandru {Paler}.
\newblock {Efficient Quantum Circuit Design with a Standard Cell Approach}.
\newblock {\em arXiv e-prints}, page arXiv:2206.04990, June 2022.

\bibitem{Wei2022}
K.~X. Wei, E.~Magesan, I.~Lauer, S.~Srinivasan, D.~F. Bogorin, S.~Carnevale,
  G.~A. Keefe, Y.~Kim, D.~Klaus, W.~Landers, N.~Sundaresan, C.~Wang, E.~J.
  Zhang, M.~Steffen, O.~E. Dial, D.~C. McKay, and A.~Kandala.
\newblock Hamiltonian engineering with multicolor drives for fast entangling
  gates and quantum crosstalk cancellation.
\newblock {\em Phys. Rev. Lett.}, 129:060501, Aug 2022.

\bibitem{Dogan2022}
Ebru Dogan, Dario Rosenstock, Loïck~Le Guevel, Haonan Xiong, Raymond~A.
  Mencia, Aaron Somoroff, Konstantin~N. Nesterov, Maxim~G. Vavilov, Vladimir~E.
  Manucharyan, and Chen Wang.
\newblock Demonstration of the two-fluxonium cross-resonance gate, 2022.

\bibitem{ding2023highfidelity}
Leon Ding, Max Hays, Youngkyu Sung, Bharath Kannan, Junyoung An, Agustin~Di
  Paolo, Amir~H. Karamlou, Thomas~M. Hazard, Kate Azar, David~K. Kim,
  Bethany~M. Niedzielski, Alexander Melville, Mollie~E. Schwartz, Jonilyn~L.
  Yoder, Terry~P. Orlando, Simon Gustavsson, Jeffrey~A. Grover, Kyle Serniak,
  and William~D. Oliver.
\newblock High-fidelity, frequency-flexible two-qubit fluxonium gates with a
  transmon coupler, 2023.

\bibitem{Milul2023}
Ofir Milul, Barkay Guttel, Uri Goldblatt, Sergey Hazanov, Lalit~M. Joshi,
  Daniel Chausovsky, Nitzan Kahn, Engin Çiftyürek, Fabien Lafont, and Serge
  Rosenblum.
\newblock A superconducting quantum memory with tens of milliseconds coherence
  time, 2023.

\bibitem{Burkhart2021}
Luke~D. Burkhart, James~D. Teoh, Yaxing Zhang, Christopher~J. Axline, Luigi
  Frunzio, M.~H. Devoret, Liang Jiang, S.~M. Girvin, and R.~J. Schoelkopf.
\newblock Error-detected state transfer and entanglement in a superconducting
  quantum network.
\newblock {\em PRX Quantum}, 2, 2021.

\bibitem{Leung2019}
N.~Leung, Y.~Lu, S.~Chakram, R.~K. Naik, N.~Earnest, R.~Ma, K.~Jacobs, A.~N.
  Cleland, and D.~I. Schuster.
\newblock Deterministic bidirectional communication and remote entanglement
  generation between superconducting qubits.
\newblock {\em npj Quantum Information}, 5, 2019.

\bibitem{Gao2021}
Yvonne~Y. Gao, M.~Adriaan Rol, Steven Touzard, and Chen Wang.
\newblock Practical guide for building superconducting quantum devices.
\newblock {\em PRX Quantum}, 2, 2021.

\bibitem{Thaker2006}
Darshan~D. Thaker, Tzvetan~S. Metodi, Andrew~W. Cross, Isaac~L. Chuang, and
  Frederic~T. Chong.
\newblock Quantum memory hierarchies: Efficient designs to match available
  parallelism in quantum computing.
\newblock volume 2006, 2006.

\bibitem{Nguyen2019}
Long~B. Nguyen, Yen-Hsiang Lin, Aaron Somoroff, Raymond Mencia, Nicholas
  Grabon, and Vladimir~E. Manucharyan.
\newblock High-coherence fluxonium qubit.
\newblock {\em Phys. Rev. X}, 9:041041, Nov 2019.

\bibitem{Krinner2022}
Sebastian Krinner, Nathan Lacroix, Ants Remm, Agustin {Di Paolo}, Elie Genois,
  Catherine Leroux, Christoph Hellings, Stefania Lazar, Francois Swiadek,
  Johannes Herrmann, Graham~J Norris, Christian~Kraglund Andersen, Markus
  M{\"{u}}ller, Alexandre Blais, Christopher Eichler, and Andreas Wallraff.
\newblock {Realizing repeated quantum error correction in a distance-three
  surface code}.
\newblock {\em Nature}, 605(7911):669--674, 2022.

\bibitem{Acharya2023}
Rajeev Acharya, Igor Aleiner, Richard Allen, Trond~I. Andersen, Markus Ansmann,
  Frank Arute, Kunal Arya, Abraham Asfaw, Juan Atalaya, Ryan Babbush, Dave
  Bacon, Joseph~C. Bardin, Joao Basso, Andreas Bengtsson, Sergio Boixo, Gina
  Bortoli, Alexandre Bourassa, Jenna Bovaird, Leon Brill, Michael Broughton,
  Bob~B. Buckley, David~A. Buell, Tim Burger, Brian Burkett, Nicholas Bushnell,
  Yu~Chen, Zijun Chen, Ben Chiaro, Josh Cogan, Roberto Collins, Paul Conner,
  William Courtney, Alexander~L. Crook, Ben Curtin, Dripto~M. Debroy, Alexander
  Del~Toro Barba, Sean Demura, Andrew Dunsworth, Daniel Eppens, Catherine
  Erickson, Lara Faoro, Edward Farhi, Reza Fatemi, Leslie~Flores Burgos,
  Ebrahim Forati, Austin~G. Fowler, Brooks Foxen, William Giang, Craig Gidney,
  Dar Gilboa, Marissa Giustina, Alejandro~Grajales Dau, Jonathan~A. Gross,
  Steve Habegger, Michael~C. Hamilton, Matthew~P. Harrigan, Sean~D. Harrington,
  Oscar Higgott, Jeremy Hilton, Markus Hoffmann, Sabrina Hong, Trent Huang,
  Ashley Huff, William~J. Huggins, Lev~B. Ioffe, Sergei~V. Isakov, Justin
  Iveland, Evan Jeffrey, Zhang Jiang, Cody Jones, Pavol Juhas, Dvir Kafri,
  Kostyantyn Kechedzhi, Julian Kelly, Tanuj Khattar, Mostafa Khezri, Mária
  Kieferová, Seon Kim, Alexei Kitaev, Paul~V. Klimov, Andrey~R. Klots,
  Alexander~N. Korotkov, Fedor Kostritsa, John~Mark Kreikebaum, David Landhuis,
  Pavel Laptev, Kim-Ming Lau, Lily Laws, Joonho Lee, Kenny Lee, Brian~J.
  Lester, Alexander Lill, Wayne Liu, Aditya Locharla, Erik Lucero, Fionn~D.
  Malone, Jeffrey Marshall, Orion Martin, Jarrod~R. McClean, Trevor Mccourt,
  Matt McEwen, Anthony Megrant, Bernardo~Meurer Costa, Xiao Mi, Kevin~C. Miao,
  Masoud Mohseni, Shirin Montazeri, Alexis Morvan, Emily Mount, Wojciech
  Mruczkiewicz, Ofer Naaman, Matthew Neeley, Charles Neill, Ani Nersisyan,
  Hartmut Neven, Michael Newman, Jiun~How Ng, Anthony Nguyen, Murray Nguyen,
  Murphy~Yuezhen Niu, Thomas~E. O'Brien, Alex Opremcak, John Platt, Andre
  Petukhov, Rebecca Potter, Leonid~P. Pryadko, Chris Quintana, Pedram Roushan,
  Nicholas~C. Rubin, Negar Saei, Daniel Sank, Kannan Sankaragomathi, Kevin~J.
  Satzinger, Henry~F. Schurkus, Christopher Schuster, Michael~J. Shearn, Aaron
  Shorter, Vladimir Shvarts, Jindra Skruzny, Vadim Smelyanskiy, W.~Clarke
  Smith, George Sterling, Doug Strain, Marco Szalay, Alfredo Torres, Guifre
  Vidal, Benjamin Villalonga, Catherine~Vollgraff Heidweiller, Theodore White,
  Cheng Xing, Z.~Jamie Yao, Ping Yeh, Juhwan Yoo, Grayson Young, Adam Zalcman,
  Yaxing Zhang, and Ningfeng Zhu.
\newblock Suppressing quantum errors by scaling a surface code logical qubit.
\newblock {\em Nature}, 614:676–681, 2023.

\bibitem{Read2022}
Alexander~P. Read, Benjamin~J. Chapman, Chan~U Lei, Jacob~C. Curtis, Suhas
  Ganjam, Lev Krayzman, Luigi Frunzio, and Robert~J. Schoelkopf.
\newblock Precision measurement of the microwave dielectric loss of sapphire in
  the quantum regime with parts-per-billion sensitivity.
\newblock 2022.

\bibitem{Place2022}
Alexander Patrick~McCormick Place.
\newblock {\em Increasing Lifetimes of Superconducting Qubits}.
\newblock PhD thesis, Princeton University, 2022.

\bibitem{PhysRevApplied.16.024023}
H.J. Mamin, E.~Huang, S.~Carnevale, C.T. Rettner, N.~Arellano, M.H. Sherwood,
  C.~Kurter, B.~Trimm, M.~Sandberg, R.M. Shelby, M.A. Mueed, B.A. Madon,
  A.~Pushp, M.~Steffen, and D.~Rugar.
\newblock Merged-element transmons: Design and qubit performance.
\newblock {\em Phys. Rev. Appl.}, 16:024023, Aug 2021.

\bibitem{Mamin2022MM}
Harry~J Mamin.
\newblock The merged element transmon.
\newblock {\em Bulletin of the American Physical Society}, 2022.

\bibitem{Mariantoni2011}
Matteo Mariantoni, H.~Wang, T.~Yamamoto, M.~Neeley, Radoslaw~C. Bialczak,
  Y.~Chen, M.~Lenander, Erik Lucero, A.~D. O'Connell, D.~Sank, M.~Weides,
  J.~Wenner, Y.~Yin, J.~Zhao, A.~N. Korotkov, A.~N. Cleland, and John~M.
  Martinis.
\newblock Implementing the quantum von neumann architecture with
  superconducting circuits.
\newblock {\em Science}, 334, 2011.

\bibitem{Reagor2016}
Matthew Reagor, Wolfgang Pfaff, Christopher Axline, Reinier~W. Heeres, Nissim
  Ofek, Katrina Sliwa, Eric Holland, Chen Wang, Jacob Blumoff, Kevin Chou,
  Michael~J. Hatridge, Luigi Frunzio, Michel~H. Devoret, Liang Jiang, and
  Robert~J. Schoelkopf.
\newblock Quantum memory with millisecond coherence in circuit qed.
\newblock {\em Phys. Rev. B}, 94:014506, Jul 2016.

\bibitem{Matanin2023}
Aleksei~R. Matanin, Konstantin~I. Gerasimov, Eugene~S. Moiseev, Nikita~S.
  Smirnov, Anton~I. Ivanov, Elizaveta~I. Malevannaya, Victor~I. Polozov,
  Eugeny~V. Zikiy, Andrey~A. Samoilov, Ilya~A. Rodionov, and Sergey~A. Moiseev.
\newblock Toward highly efficient multimode superconducting quantum memory.
\newblock {\em Phys. Rev. Appl.}, 19:034011, Mar 2023.

\bibitem{Axline2016}
C.~Axline, M.~Reagor, R.~Heeres, P.~Reinhold, C.~Wang, K.~Shain, W.~Pfaff,
  Y.~Chu, L.~Frunzio, and R.~J. Schoelkopf.
\newblock An architecture for integrating planar and 3d cqed devices.
\newblock {\em Applied Physics Letters}, 109, 2016.

\bibitem{MacCabe2020}
Gregory~S. MacCabe, Hengjiang Ren, Jie Luo, Justin~D. Cohen, Hengyun Zhou, Alp
  Sipahigil, Mohammad Mirhosseini, and Oskar Painter.
\newblock Nano-acoustic resonator with ultralong phonon lifetime.
\newblock {\em Science}, 370, 2020.

\bibitem{Pechal2019}
Marek Pechal, Patricio Arrangoiz-Arriola, and Amir~H. Safavi-Naeini.
\newblock Superconducting circuit quantum computing with nanomechanical
  resonators as storage.
\newblock {\em Quantum Science and Technology}, 4, 2019.

\bibitem{Krinner2019}
S.~Krinner, S.~Storz, P.~Kurpiers, P.~Magnard, J.~Heinsoo, R.~Keller,
  J.~Lütolf, C.~Eichler, and A.~Wallraff.
\newblock Engineering cryogenic setups for 100-qubit scale superconducting
  circuit systems.
\newblock {\em EPJ Quantum Technology}, 6, 2019.

\bibitem{Bravyi2022}
Sergey Bravyi, Oliver Dial, Jay~M. Gambetta, Darió Gil, and Zaira Nazario.
\newblock The future of quantum computing with superconducting qubits.
\newblock {\em Journal of Applied Physics}, 132, 2022.

\bibitem{Brecht2016}
Teres~A. Brecht, Wolfgang Pfaff, Chen Wang, Yiwen Chu, Luigi Frunzio, Michel~H.
  Devoret, and Robert~J. Schoelkopf.
\newblock Multilayer microwave integrated quantum circuits for scalable quantum
  computing.
\newblock {\em npj Quantum Information}, 2, 2016.

\bibitem{Kosen2022}
Sandoko Kosen, Hang~Xi Li, Marcus Rommel, Daryoush Shiri, Christopher Warren,
  Leif Grönberg, Jaakko Salonen, Tahereh Abad, Janka Biznárová, Marco
  Caputo, Liangyu Chen, Kestutis Grigoras, Göran Johansson, Anton~Frisk
  Kockum, Christian KriÅ3/4an, Daniel~Pérez Lozano, Graham~J. Norris, Amr
  Osman, Jorge Fernández-Pendás, Alberto Ronzani, Anita~Fadavi Roudsari,
  Slawomir Simbierowicz, Giovanna Tancredi, Andreas Wallraff, Christopher
  Eichler, Joonas Govenius, and Jonas Bylander.
\newblock Building blocks of a flip-chip integrated superconducting quantum
  processor.
\newblock {\em Quantum Science and Technology}, 7, 2022.

\bibitem{Sete2015}
Eyob~A. Sete, John~M. Martinis, and Alexander~N. Korotkov.
\newblock Quantum theory of a bandpass purcell filter for qubit readout.
\newblock {\em Phys. Rev. A}, 92:012325, Jul 2015.

\bibitem{Qiskit}
{Qiskit contributors}.
\newblock Qiskit: An open-source framework for quantum computing, 2023.

\bibitem{PhysRevLett.127.040503}
Stefan Krastanov, Hamza Raniwala, Jeffrey Holzgrafe, Kurt Jacobs, Marko
  Lon\ifmmode~\check{c}\else \v{c}\fi{}ar, Matthew~J. Reagor, and Dirk~R.
  Englund.
\newblock Optically heralded entanglement of superconducting systems in quantum
  networks.
\newblock {\em Phys. Rev. Lett.}, 127:040503, Jul 2021.

\bibitem{DEJMPS_1996}
David Deutsch, Artur Ekert, Richard Jozsa, Chiara Macchiavello, Sandu Popescu,
  and Anna Sanpera.
\newblock Quantum privacy amplification and the security of quantum
  cryptography over noisy channels.
\newblock {\em Physical Review Letters}, 77(13):2818--2821, sep 1996.

\bibitem{fowler2012surface}
Austin~G Fowler, Matteo Mariantoni, John~M Martinis, and Andrew~N Cleland.
\newblock Surface codes: Towards practical large-scale quantum computation.
\newblock {\em Physical Review A}, 86(3):032324, 2012.

\bibitem{2017QS&T....2c5008C}
Christopher {Chamberland} and Tomas {Jochym-O'Connor}.
\newblock {Error suppression via complementary gauge choices in Reed-Muller
  codes}.
\newblock {\em Quantum Science and Technology}, 2(3):035008, September 2017.

\bibitem{Noh2022}
Kyungjoo Noh, Christopher Chamberland, and Fernando~G.S.L. Brand\~ao.
\newblock Low-overhead fault-tolerant quantum error correction with the
  surface-gkp code.
\newblock {\em PRX Quantum}, 3:010315, Jan 2022.

\bibitem{Teoh2022}
James~D. Teoh, Patrick Winkel, Harshvardhan~K. Babla, Benjamin~J. Chapman,
  Jahan Claes, Stijn~J. de~Graaf, John W.~O. Garmon, William~D. Kalfus, Yao Lu,
  Aniket Maiti, Kaavya Sahay, Neel Thakur, Takahiro Tsunoda, Sophia~H. Xue,
  Luigi Frunzio, Steven~M. Girvin, Shruti Puri, and Robert~J. Schoelkopf.
\newblock Dual-rail encoding with superconducting cavities.
\newblock 2022.

\bibitem{StimRepo}
{Stim Github repository}.
\newblock https://github.com/quantumlib/Stim, 2023.

\bibitem{Pechal2021}
M.~Pechal, G.~Salis, M.~Ganzhorn, D.~J. Egger, M.~Werninghaus, and S.~Filipp.
\newblock Characterization and tomography of a hidden qubit.
\newblock {\em Phys. Rev. X}, 11:041032, Nov 2021.

\bibitem{2018Quant...2...53C}
Christopher {Chamberland} and Michael~E. {Beverland}.
\newblock {Flag fault-tolerant error correction with arbitrary distance codes}.
\newblock {\em Quantum}, 2:53, February 2018.

\bibitem{Gottesman:2022jlv}
Daniel Gottesman.
\newblock {Opportunities and Challenges in Fault-Tolerant Quantum Computation}.
\newblock 10 2022.

\bibitem{bravyi2015doubled}
Sergey Bravyi and Andrew Cross.
\newblock Doubled color codes.
\newblock {\em arXiv preprint arXiv:1509.03239}, 2015.

\bibitem{choi2013cost}
Byung-Soo Choi.
\newblock Cost comparison between code teleportation and stabilizer sequence
  methods for quantum code conversion.
\newblock In {\em 2013 International Conference on ICT Convergence (ICTC)},
  pages 1083--1087, 2013.

\bibitem{eastin2009restrictions}
Bryan Eastin and Emanuel Knill.
\newblock Restrictions on transversal encoded quantum gate sets.
\newblock {\em Physical review letters}, 102(11):110502, 2009.

\bibitem{zeng2011transversality}
Bei Zeng, Andrew Cross, and Isaac~L Chuang.
\newblock Transversality versus universality for additive quantum codes.
\newblock {\em IEEE Transactions on Information Theory}, 57(9):6272--6284,
  2011.

\bibitem{bravyi2012magic}
Sergey Bravyi and Jeongwan Haah.
\newblock Magic-state distillation with low overhead.
\newblock {\em Physical Review A}, 86(5):052329, 2012.

\bibitem{litinski2019magic}
Daniel Litinski.
\newblock Magic state distillation: Not as costly as you think.
\newblock {\em Quantum}, 3:205, 2019.

\bibitem{szkopek2006threshold}
Thomas Szkopek, P~Oscar Boykin, Heng Fan, Vwani~P Roychowdhury, Eli
  Yablonovitch, Geoffrey Simms, Mark Gyure, and Bryan Fong.
\newblock Threshold error penalty for fault-tolerant quantum computation with
  nearest neighbor communication.
\newblock {\em IEEE transactions on nanotechnology}, 5(1):42--49, 2006.

\bibitem{Young:2022cyz}
C.~B. Young, A.~Safari, P.~Huft, J.~Zhang, E.~Oh, R.~Chinnarasu, and
  M.~Saffman.
\newblock {An architecture for quantum networking of neutral atom processors}.
\newblock {\em Appl. Phys. B}, 128(8):151, 2022.

\bibitem{Bluvstein:2021jsq}
Dolev Bluvstein et~al.
\newblock {A quantum processor based on coherent transport of entangled atom
  arrays}.
\newblock {\em Nature}, 604(7906):451--456, 2022.

\bibitem{2019ApPRv...6b1314B}
Colin~D. {Bruzewicz}, John {Chiaverini}, Robert {McConnell}, and Jeremy~M.
  {Sage}.
\newblock {Trapped-ion quantum computing: Progress and challenges}.
\newblock {\em Applied Physics Reviews}, 6(2):021314, June 2019.

\bibitem{2021Natur.592..209P}
J.~M. {Pino}, J.~M. {Dreiling}, C.~{Figgatt}, J.~P. {Gaebler}, S.~A. {Moses},
  M.~S. {Allman}, C.~H. {Baldwin}, M.~{Foss-Feig}, D.~{Hayes}, K.~{Mayer},
  C.~{Ryan-Anderson}, and B.~{Neyenhuis}.
\newblock {Demonstration of the trapped-ion quantum CCD computer architecture}.
\newblock {\em \nat}, 592(7853):209--213, January 2021.

\bibitem{2020AVSQS...2a4101K}
V.~{Kaushal}, B.~{Lekitsch}, A.~{Stahl}, J.~{Hilder}, D.~{Pijn},
  C.~{Schmiegelow}, A.~{Bermudez}, M.~{M{\"u}ller}, F.~{Schmidt-Kaler}, and
  U.~{Poschinger}.
\newblock {Shuttling-based trapped-ion quantum information processing}.
\newblock {\em AVS Quantum Science}, 2(1):014101, February 2020.

\bibitem{2016NatCo...710352S}
C.~{Schuck}, X.~{Guo}, L.~{Fan}, X.~{Ma}, M.~{Poot}, and H.~X. {Tang}.
\newblock {Quantum interference in heterogeneous superconducting-photonic
  circuits on a silicon chip}.
\newblock {\em Nature Communications}, 7:10352, January 2016.

\bibitem{2021npjQI...7..121N}
Tom{\'a}{\v{s}} {Neuman}, Matt {Eichenfield}, Matthew~E. {Trusheim}, Lisa
  {Hackett}, Prineha {Narang}, and Dirk {Englund}.
\newblock {A phononic interface between a superconducting quantum processor and
  quantum networked spin memories}.
\newblock {\em npj Quantum Information}, 7:121, January 2021.

\bibitem{Zhou2021}
Chao Zhou, Pinlei Lu, Matthieu Praquin, Tzu-Chiao Chien, Ryan Kaufman, Xi~Cao,
  Mingkang Xia, Roger Mong, Wolfgang Pfaff, David Pekker, and Michael Hatridge.
\newblock A modular quantum computer based on a quantum state router.
\newblock 2021.

\bibitem{kang2014cmos}
S.M. Kang, Y.~Leblebici, and C.W. Kim.
\newblock {\em CMOS Digital Integrated Circuits Analysis \& Design}.
\newblock McGraw-Hill Education, 2014.

\bibitem{1540969}
T.S. Metodi, D.D. Thaker, A.W. Cross, F.T. Chong, and I.L. Chuang.
\newblock A quantum logic array microarchitecture: scalable quantum data
  movement and computation.
\newblock In {\em 38th Annual IEEE/ACM International Symposium on
  Microarchitecture (MICRO'05)}, pages 12 pp.--318, 2005.

\end{thebibliography}
